\documentclass[pre,twocolumn,showpacs,superscriptaddress,preprintnumbers,floatfix]{revtex4}

\usepackage{dcolumn}
\usepackage{amsmath}
\usepackage{amssymb}
\usepackage{graphicx}
\usepackage{bm}   
\usepackage{bbm}   
\usepackage{verbatim}
\usepackage{stmaryrd}
\usepackage{amsthm}

\def\clap#1{\hbox to 0pt{\hss#1\hss}}

\def\mathclap{\mathpalette\mathclapinternal}

\def\mathclapinternal#1#2{%
  \clap{$\mathsurround=0pt#1{#2}$}}
\def\bra#1{\mathinner{\langle{#1}|}}
\def\ket#1{\mathinner{|{#1}\rangle}}
\def\braket#1{\mathinner{\langle{#1}\rangle}}

\theoremstyle{break} 	\newtheorem{Cor}{Corollary}
\theoremstyle{plain} 	\newtheorem*{ProCor}{Proof}
\theoremstyle{break} 	\newtheorem{The}{Theorem}
\theoremstyle{plain} 	\newtheorem*{ProThe}{Proof}
\theoremstyle{break} 	\newtheorem{Prop}{Proposition}
\theoremstyle{plain} 	\newtheorem*{ProProp}{Proof}
\theoremstyle{break} 	
\theoremstyle{plain}	
\theoremstyle{break}	\newtheorem*{Def}{Definition} 
\theoremstyle{plain}	\newtheorem*{Not}{Notation}

\newcommand{\Prob}      {\mathrm{Pr}}
\newcommand{\Lang}      {\mathcal{L}}
\newcommand{\Graph}      {\mathcal{G}}
\newcommand{\supp}      {\mathrm{supp}\;}

\newcommand{\prcl}      {\mathcal{P}}
\newcommand{\PRCL}      {\mathcal{\mathbf{P}}} 

\newcommand{\stochL}    {\mathcal{S}}

\newcommand{\field}[1]{\mathbb{#1}}
\newcommand{\rank}      {\mathrm{dim}}
\newcommand{\symb}      {\sigma}

\newcommand{\mperiod} {\mathnormal{p}}

\newcommand{\mach}      {\mathcal{M}}

\newcommand{\dimn}       {\mathrm{dim}}
\newcommand{\tr}        {\mathrm{tr}}

\begin{document}


\title{Computation in Finitary Stochastic and Quantum Processes}

\author{Karoline Wiesner}
\email{k.wiesner@bristol.ac.uk}
\affiliation{Complexity Sciences Center and Physics Department,
University of California Davis, One Shields Avenue, Davis, CA 95616}
\affiliation{Santa Fe Institute, 1399 Hyde Park Road, Santa Fe, NM 87501}
\author{James P. Crutchfield}
\email{chaos@cse.ucdavis.edu}
\affiliation{Complexity Sciences Center and Physics Department,
University of California Davis, One Shields Avenue, Davis, CA 95616}
\affiliation{Santa Fe Institute, 1399 Hyde Park Road, Santa Fe, NM 87501}

\date{\today}

\bibliographystyle{unsrt}

\begin{abstract}
We introduce \emph{stochastic} and \emph{quantum finite-state transducers}
as computation-theoretic models of classical stochastic and quantum finitary
processes. Formal \emph{process languages}, representing the distribution
over a process's behaviors, are recognized and generated by suitable
specializations. We characterize and compare deterministic and nondeterministic
versions, summarizing their relative computational power in a hierarchy of
finitary process languages. Quantum finite-state transducers and generators
are a first step toward a computation-theoretic analysis of individual,
repeatedly measured quantum dynamical systems. They are explored via several
physical systems, including an iterated beam splitter, an atom in a magnetic
field, and atoms in an ion trap---a special case of which implements the
Deutsch quantum algorithm. We show that these systems' behaviors, and so their
information processing capacity, depends sensitively on the measurement
protocol.

\end{abstract}

\preprint{Santa Fe Institute Working Paper 06-09-031}
\preprint{arxiv.org/quant-ph/0608206}

\maketitle


\tableofcontents


\section{Introduction}

Automata theory is the study of abstract computing devices, or \emph{machines},
and the class of functions they can perform on their inputs. In the 1940's
and 1950's, simple kinds of machines, so-called \emph{finite-state automata},
were introduced to model brain function \cite{mccu43a,arbi02a}. They turned
out to be extremely useful for a variety of other purposes, such as studying
the lower limits of computational power and synthesizing logic controllers
and communication networks. In the late 1950's, the linguist Noam Chomsky
developed a classification of \emph{formal languages} in terms of the
\emph{grammars} and automata required to recognize them \cite{chomsky:56}.
On the lowest level of Chomsky's hierarchy, for example, whether or not a
given sentence obeys the grammatical rules of a language is answered by a
finite-state automaton.

Our understanding of the nature of computing has changed substantially in
the intervening half century. In recent years the study of computation
with elementary components that obey quantum mechanical laws has developed
into a highly active research area. 

\subsection{Finite Quantum Computing}

The physical laws underlying quantum computation are a mixed blessing.
On the one hand, a growing body of theoretical results suggests that a
computational device whose components are directly governed by quantum
physics may be considerably more powerful than its classical counterpart.
Undoubtedly, the most celebrated of these results is Shor's factoring
algorithm from 1994 \cite{shor:94}. Other results include Grover's
quantum search algorithm from 1996 \cite{grover:96}. On the other hand,
the results employ powerful computational architectures, such as quantum Turing
machines \cite{deutsch:85}, that are decidedly more powerful than finite-state
machines and that must maintain high degrees of internal coherence and
environmental isolation during operation. For a review of theoretical and
experimental studies of quantum computation see, for example, Refs.
\cite{galindo:02,zoll05}.

However, to date, implementation efforts have fallen substantially short of the
theoretical promise. So far experimental tests of quantum computation are on
small-scale systems---in fact, very small. Currently, the largest coherent
system of information storage supports only $7$ quantum bits or \emph{qubits}
\cite{knill:00}. Thus, the study of finite-state quantum automata
is motivated by very practical concerns. They reflect the capabilities of currently
feasible quantum computers. As was also true in the first days
of digital computers, though, the study of finite machines is also a
starting point, here for developing a computational hierarchy for quantum
dynamical systems.

\subsection{Dynamics, Information, and Measurement}

A common goal in the practice of quantum theory is to predict the expectation
of outcomes from an ensemble of isolated measurements. There is a key
difference, though, between this and what one needs to understand quantum
processes. For quantum processes, such as found in molecular dynamics, one
must analyze \emph{behavior}; predicting an observable's mean value is
insufficient.

Quantum mechanics can be extended, of course, to address behavior. This has
been done in rather general frameworks (e.g., Ref. \cite{alicki:01}), as well
as in special cases, such as quantum Markov chains \cite{kumm02}. However,
questions about a quantum system's information processing capacity remain
unanswered.
For example, how much of a quantum system's history is stored in its state?
How is that information processed to produce future behavior? More pointedly,
even if a system is designed to have a desired information processing capacity,
the question always remains whether or not that capacity is actually used
during operation.

An intriguing, but seemingly unrelated area of research in quantum behavior
is \emph{quantum chaos}---the production of information through the
exponential amplification of perturbations \cite{gutzwiller}.
Since any quantum system is described by the Schr\"odinger
equation, which is linear, chaotic behavior cannot arise. However, quantum
systems that exhibit chaotic behavior in the classical limit can show
signatures of chaos in semi-classical regimes. Thus, analyzing the
relationship between classical and quantum dynamical systems plays an
important role in understanding the origins of quantum information
production.

For quantum systems, in contrast with their classical counterparts,
including measurement interactions is essential to any complete description.
Unfortunately, this is largely missing from current dynamical theories.
Nonetheless, simulation studies show that measurement interactions lead to
genuinely chaotic behavior in quantum dynamical systems, even far from the
semi-classical limit \cite{habi06}. Observation must be the basis for modeling
a quantum process---either in describing its behavior or quantifying its
computational capacity.

\subsection{Technical Setting}

Here we introduce finite computation-theoretic quantum models that, when
analyzed with tools from quantum mechanics and stochastic processes,
simultaneously embody dynamics, measurement, and information processing.
Studies of quantum chaos are, in effect, extensions of the theory of nonlinear
(classical) dynamics. Dynamical systems are often analyzed by transforming
them into finite-state automata using the methods of \emph{symbolic dynamics}
\cite{lind}. The quantum automata in the following model dynamical behavior
and include measurement interactions and so provide a kind of symbolic
dynamics for quantum systems \cite{Wies07a}. The result is a line of inquiry
complementary to both quantum computation and quantum dynamical systems.

One goal is to develop a representation of quantum processes that allows one
to analyze their intrinsic computation. \emph{Intrinsic computation} in a
dynamical system is an inherent property of the behavior the system generates
\cite{crut94}. One asks three basic questions of the system: First,
how much historical information is stored in the current state? Second, in
what architecture is that information stored? Finally, how is the stored
information transformed to produce future behavior? This approach has been
used to analyze intrinsic computation in classical dynamical systems,
statistical mechanical systems, and stochastic processes
\cite{crutchfield:89,crutchfield:90-1,crutchfield:93,crutchfield:97}.

We view the present contribution as a direct extension of this prior work
and, also, as complementary to the current design and theoretical-engineering
approach to quantum computation. Specifically, we focus on the dynamics of
quantum processes, rather than on methods to construct devices that implement
a desired function. We express the intrinsic information processing using
various kinds of finite-memory devices. We emphasize the effects of measurement
on a quantum system's behavior and so, in this way, provide a somewhat
different view of quantum dynamical systems for which, typically, observation
is ignored. An information-theoretic analysis using the resulting framework
can be found in Refs.~\cite{Crut06a,Wies07a}.

Most directly, we are interested, as natural scientists are, in
\emph{behavior}---how a system state develops over time. In the
computation-theoretic setting this translates into a need to model
\emph{generators}. In contrast, the conventional setting for analyzing
the computational power of automata centers around detecting membership
of words in a language. As a consequence, the overwhelming fraction of
existing results on automata concerns devices that \emph{recognize} an input
string---and on problems that can be recast as such.
Automata that spontaneously generate outputs are much less often
encountered, if at all, in the theory of computation. Nonetheless, generators
are necessary if one wants to model physical processes using dynamical systems.
In particular, as we hope to show, quantum generators are a key tool for
detecting information processing capabilities inherent in natural processes.

\subsection{Overview}

Due to the range of topics, in the following we give a selective, but
self-contained treatment. We review what is needed from automata, formal
languages, and quantum theory, though familiarity with those areas is helpful. 
Citations to reference texts are given at the appropriate points.

Our approach will make most sense, especially to those unfamiliar with the
theory of formal languages, if we devote some time to reviewing basic
automata theory and its original goals. This also allows us to establish,
in a graded fashion, the necessary notation for the full development,
clearly identifying which properties are quantum mechanical and which,
in contrast, are essentially classical (and probabilistic). In
addition, this illustrates one of the principle benefits of discrete
computation theory: i.e., the classification of devices that
implement different kinds of information processing. Those for whom
automata and formal languages are well known, though, should appreciate
by the end of the review the physical and dynamical motivations, since
these will be expressed within the existing frameworks of discrete
computation and stochastic processes.

To lay the foundations for a computational perspective on quantum dynamical
systems the most basic notion we introduce is the class of finite-state automata
called \emph{quantum finite-state transducers}. To get to these, in the next
sections we introduce the concept of
\emph{process languages}, building on formal language theory. We then present
stochastic finite-state transducers and their subclasses---stochastic
recognizers and generators---as classical representations of process languages. 
The relationship between automata and languages is discussed in each case and
we provide an overview (and introduce notation) that anticipates their quantum
analogs. We then introduce quantum finite-state transducers and their
subclasses---quantum recognizers and generators---and discuss their various
properties. Finally, we illustrate the main ideas by analyzing specific
examples of quantum dynamical systems that they can model.


\section{Finitary Stochastic Processes}
\label{sec:FinStocProc}

Consider the temporal evolution of the state of some natural system. The
evolution is monitored by a series of measurements---numbers registered in
some way, perhaps continuously, perhaps discretely. Each such measurement
can be taken as a random variable. The distribution over sequences of these
random variables is what we refer to as a \emph{stochastic process}. An
important question for understanding the structure of natural systems is
what kinds of stochastic processes there are.

The class of \emph{finitary} stochastic processes was introduced to identify
those that require only a finite amount of internal resources to generate
their behavior. This property is important in several settings. In symbolic
dynamical systems, for example, it was shown that the \emph{sofic subshifts}
have a form of infinite correlation in their temporal behaviors despite being
finitely specified \cite{weis73}. The information-theoretic characterization of
stochastic processes \cite{bial01,vita02,crut03}, as another example, defines finitary
processes as those with a bounded value of mutual information between past and
future behaviors. Here, we remain close to these original definitions, giving
explicit structural models, both classical and quantum, for finitary processes.

In this, we use formal language theory.
Our use of formal language theory differs from most, though, in how it
analyzes the connection between a language and the systems that can generate
it. In brief, we observe a system through a finite-resolution measuring
instrument, representing each output with a \emph{symbol} $\symb$ from
discrete \emph{alphabet} $\Sigma$. The temporal behavior of a system, then, is
a string or a \emph{word} consisting of a succession of measurement symbols.
The collection of all (and only) those words is the \emph{language}
that captures the possible, temporal behaviors of the system.

\begin{Def}
A \emph{formal language} $\mathcal{L}$ is a set of \emph{words}
$w = \symb_0\symb_1\symb _2  \ldots $ each of which consists of a finite 
series of symbols $\symb_t \in \Sigma$ from a discrete alphabet $\Sigma$. 
\end{Def}

In the following $\lambda$ denotes the empty word. $\Sigma^*$ denotes the set
of all possible
words, including $\lambda$, of any length formed using symbols in $\Sigma$.
We denote a word of length $L$ by $\symb^L = \symb_0 \symb_1 \ldots
\symb_{L-1}$, with $\symb_t \in \Sigma$. The set of all words of length $L$ is
$\Sigma^L$.

Since a formal language, as we use the term, is a set of observed words
generated by a process, then each \emph{subword} $\symb_t \symb_{t+1} \ldots
\symb_{u-1} \symb_u$, $t \leq u$, $t,u = 0,1, \ldots, L-1$, of a word
$\symb^L$ has also been observed and is considered part of the language. This
leads to the following definition.

\begin{Def}
A language $\mathcal{L}$ is \emph{subword closed} if, for each
$w \in \Lang$, all of $w$'s subwords $\mathrm{sub}(w)$ are also
members of $\Lang$: $\mathrm{sub}(w) \subseteq \mathcal{L}$.
\end{Def}

Finally, we imagine that a system can run for an arbitrarily
long time and so the language describing its behaviors has words of
arbitrary length. In this way, a subword-closed formal language---as
a set of arbitrarily long series of measurement outcomes---represents the
allowed (and, implicitly, disallowed) behaviors of a system.

Beyond a formal language listing which words (or behaviors) occur and
which do not, we are also interested in the probability of their
occurrence. Let $\Prob(w)$ denote the probability of word $w$, then
we have the following.

\begin{Def}
A \emph{stochastic language} $\stochL$ is a formal language with a
\emph{word distribution} $\Prob(w)$ that is normalized at each length $L$:
\begin{equation}
\sum_{\mathclap{\{w \in \Sigma^L \}}} \Prob(w) = 1 ~, L = 1, 2, 3, \ldots
\end{equation}
with $0 \leq \Prob(w) \leq 1$ .
\end{Def}

\begin{Def}
The \emph{joint probability} of symbol $\symb$ following word $w$ 
is written $\Prob(w\symb)$. 
\end{Def}

\begin{Def}
The \emph{conditional probability} $\Prob(\symb|w)$ of symbol $\symb$ given the
preceding observation of word $w$ is 
\begin{equation}
\label{eqn:cndprb}
\Prob(\symb|w) = \Prob(w\symb) / \Prob(w)~,
\end{equation}
when $\Prob(w) > 0$; otherwise, $\Prob(\symb|w) = 0$.
\end{Def}

For purposes of comparison between various computational models, it is
helpful to refer directly to the set of words in a stochastic language
$\stochL$. This is the \emph{support} of a stochastic language:
\begin{equation}
\supp(\stochL) = \{w \in \stochL: ~\Prob(w) > 0 \} ~.
\end{equation}

These lead us, finally, to define the main object of study.

\begin{Def}
A \emph{process language} ${\prcl}$ is a stochastic language that is
subword closed and it obeys the consistency condition $\Prob(\symb^L)
\geq \Prob(\symb^L\symb)$.
\end{Def}

A process language represents all of a system's possible behaviors,
$w \in \supp(\prcl)$, and their probabilities $\Prob(w)$
of occurrence. In its completeness it could
be taken as a model of the system, but at best it is a rather prosaic
and unwieldy representation. Indeed, a \emph{model} of a process is
usually intended to be a more compact description than a literal
listing of observations. In the best of circumstances a model's
components capture some aspect of a system's structure and
organization. Here we will be even more specific, the models that
we will focus on not only have to describe a process language, but
they will also consist of two structural components: states and
transitions between them. (One should contrast the seeming
obviousness of the latter with the fact that there are alternative
computational models, such as grammars, which do not use
the concept of state.)

\begin{figure}  
\begin{center}
\resizebox{!}{2.50in}{\includegraphics{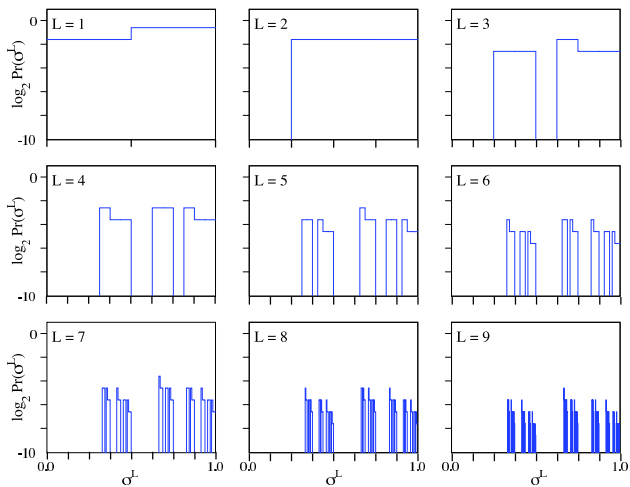}}
\end{center}
\caption{Example of a process language: In the Golden
  Mean Process, with alphabet $\Sigma = \{0,1\}$, word $00$ and all words
  containing it have zero probability. All other words have nonzero
  probability. The logarithm base 2 of the word probabilities is plotted
  versus the binary string $\symb^L$, represented as base-$2$ real number
  ``$0.\symb^L$''. To allow word probabilities to be compared at different
  lengths, the distribution is normalized on $[0,1]$---that is, the
  probabilities are calculated as densities.
  }
\label{fig:stol}
\end{figure}

To illustrate process languages we give an example in Fig.~\ref{fig:stol},
which shows a language---from the \emph{Golden Mean Process}---and its
word distribution at different word lengths. In this process language
$\Sigma = \{0,1\}$ and word $00$ and all words containing it have zero
probability. Moreover, if a $1$ is seen, then the next $\symb \in \Sigma$
occurs with fair probability.

Figure~\ref{fig:stol} plots the base-$2$
logarithm of the word probabilities versus the binary string $\symb^L$,
represented as the base-$2$ real number
$0.\symb^L  = \sum_{t=0}^{L-1}\symb_t 2^{-t-1} \in [0,1]$. At length
$L=1$ (upper leftmost plot) both words $0$ 
and $1$ are allowed but have different probabilities. At $L=2$ the first
disallowed string $00$ occurs. As $L$ grows an increasing number of
words are forbidden---those containing the shorter forbidden word $00$. As
$L \rightarrow \infty$ the set of allowed words forms a self-similar,
uncountable, closed, and disconnected (Cantor) set in the interval
$[0,1]$ \cite{lind}. Note that the language is subword closed. The process's
name comes from the fact that the logarithm of the number of allowed words
grows exponentially with $L$ at a rate given by the logarithm of the golden
mean $\phi = \tfrac{1}{2} (1+\sqrt{5})$.  

\section{Stochastic Transducers}
\label{sec:sfm}

The process languages developed above require a new kind of finite-state
machine to represent them. And so, our immediate goal is to construct
a consistent formalism for machines that can recognize, generate, and
transform process languages. We refer to the most general ones
as \emph{stochastic transducers}. We will then specialize these
transducers into recognizers and generators.

A few comments on various kinds of stochastic transducer introduced
by others will help to motivate our approach, which has the distinct
goal of representing process languages. Paz defines \emph{stochastic
sequential machines} that are, in effect, transducers \cite{paz}. Rabin
defines \emph{probabilistic automata} that are stochastic sequential machines
with no output \cite{rabi63}. Neither, though, considers process
languages or the ``generation'' of any language for that matter. Vidal et al
define \emph{stochastic transducers}, though based on a different definition
of stochastic language \cite{vida05a}. As a result, their stochastic
transducers cannot represent process languages.
 
\subsection{Definition}
\label{sec:STDefn}

Our definition of a \emph{stochastic transducer} parallels Paz's
stochastic sequential machines. 

\begin{Def}
\label{def:mea}
A \emph{stochastic finite-state transducer} (ST) is a tuple
$\{ S,X,Y, \{ T(y|x) \} \}$ where
\begin{enumerate}
\setlength{\topsep}{0pt}
\setlength{\itemsep}{0pt}
\item $S$ is a finite set of states, including a start state $s_0$.
\item $X$ and $Y$ are finite alphabets of input and output symbols,
	respectively.
\item $\{T(y|x): x \in X, y \in Y \}$ is a set of square substochastic matrices
	of $\dimn~|S|$, one for each output-input pair $y|x$. The matrix entry
	$T_{ij}(y|x)$ is the conditional probability, when in state $i$ and
	reading in symbol $x$, of going to state $j$ and emitting symbol $y$.
\end{enumerate}
\end{Def}

Generally, a stochastic transducer (ST) operates by reading in symbols that, along with the
current state, determine the next state(s) and output symbol(s). 
At each step a symbol $x \in X$ is read from the input word. The transducer
stochastically chooses a transition $T_{ij}(y|x)>0$, emits symbol $y \in Y$,
and updates its state from $i$ to $j$. An ST thus maps an input word to one
or more output words. Unless otherwise explicitly stated, in our models there
is no delay between reading an input symbol and producing the associated
output symbols.

STs are our most general model of finitary (and nonquantum) computation.
They are structured so that specialization leads to a graded family of
models of increasing sophistication.

\subsection{Graphical Representation}
\label{sec:graph}

The set $\{ T(y|x) \}$ can be represented as a
directed graph $\Graph(T)$ with the nodes corresponding to states---the matrix
row and column indices. An edge connects nodes $i$ and $j$ and corresponds to
an element $T_{ij} > 0$ that gives the nonzero transition probability from state
$i$ to state $j$. Edges are labeled $x|p|y$ with the input symbol
$x \in X$, output symbol $y \in Y$, and transition probability $p =
T_{ij}(y|x)$. 
Since an ST associates outputs with transitions, in fact, what we have
defined is a \emph{Mealy ST}, which differs from the alternative
\emph{Moore ST} in which an output is associated with a state
\cite{paz}.  

\begin{Def}
A \emph{path} is a series of edges visited sequentially when making
state-to-state transitions with $T_{ij} > 0$.
\end{Def}

\begin{Def}
A directed graph $\Graph$ is \emph{connected} if there is at least one
path between every pair of states.
\end{Def}

\begin{Def}
A directed graph $\Graph$ is \emph{strongly connected} if for every pair of
states, $i$ and $j$, there is at least one path from
$i$ to $j$ and at least one from $j$ to $i$.
\end{Def}

The states in the graph of an ST  can be classified as follows, refining the
definitions given by Paz \cite[p. 85]{paz}.

\begin{Def}
A state $j$ is a \emph{consequent} of state $i$ if there is a path beginning
at $i$ and ending at $j$.
\end{Def}

\begin{Def}
A state is called \emph{transient } if it has a consequent of which it is not
itself a consequent.
\end{Def}

\begin{Def}
A state is called \emph{recurrent } if it has at least one consequent of
which it is itself a consequent.
\end{Def}

Note that transient and recurrent states can be overlapping sets. We
therefore make the following distinctions.

\begin{Def}
A state is called \emph{asymptotically recurrent } if it is
recurrent, but not transient. 
\end{Def}

\begin{Def}
A state is called \emph{transient recurrent } if it is transient and
recurrent. 
\end{Def}

Generally speaking, an ST starts in a set of transient states and ultimately
transits to one or another of the asymptotically recurrent subsets. That
is, there can be more than one set of asymptotically recurrent states.
Unless stated otherwise, though, in the following we will consider STs
that have only a single set of asymptotically recurrent states. 

\subsection{Word Probabilities}
\label{seq:prST}

Before discussing the process languages associated with an ST we must introduce
the matrix notation required for analysis. To facilitate comparing classical
stochastic models and their quantum analogs, we adapt Dirac's bra-ket notation:
Row vectors $\bra{\cdot}$ are called \emph{bra} vectors; and column vectors
$\ket{\cdot}$, \emph{ket} vectors.

\begin{Not}
Let $\ket{ \eta }= (1, 1, \ldots ,1, 1)^T$ denote a column vector with
$|S|$ components that are all $1$s.
\end{Not}

\begin{Not}
Let $\bra{\pi } = (\pi_0, \pi_1, \ldots, \pi_{|S|-1})$ be a row vector whose
components, $0 \leq \pi_i \leq 1$, give the \emph{probability of being in
state $i$}. The vector is normalized in probability:
$\sum_{i=0}^{|S|-1} \pi_i = 1$. The \emph{initial state distribution},
with all of the probability concentrated in the start state, is denoted 
$\bra{\pi^0 }= (1, 0, \ldots, 0)$.
\end{Not}

For a series of $L$ input symbols the action of
the corresponding ST is a product of transition matrices:
\begin{equation*}
\label{eqn:STxy}
T(y^L|x^L) = T(y_0|x_0) T(y_1|x_1)\cdots T(y_{L-1}|x_{L-1}) ~,
\end{equation*}
whose elements $T_{ij}(y^L|x^L)$ give the probability of making a transition
from state $i$ to $j$ and generating output $y^L$ when reading input $x^L$.

Starting in state distribution $\bra{\pi^0 }$, the state distribution
after reading in word $x^L$ and emitting word $y^L$ is
\begin{equation}
\bra{\pi(y^L|x^L)} = \bra{ \pi^0} T(y^L|x^L)  ~.
\end{equation}
This can then be used to compute the probability of reading out word $y^L$
conditioned on reading in word $x^L$:
\begin{equation}
\Prob(y^L|x^L) = \braket{ \pi(y^L|x^L) | \eta}  ~.
\end{equation}

\section{Stochastic Recognizers and Generators}

We are ready now to specialize this general architecture into
classes of recognizing and generating devices. In each case
we address those aspects that justify our calling them models;
viz., we can calculate various properties of the process languages
that they represent directly from the machine states and transitions,
such as the word distribution and statistical properties that
derive from it.

Generally speaking, a recognizer reads in a word and has two possible outputs
for \emph{each} symbol being read in: \emph{accept} or \emph{reject}. This
differs from the common model \cite{hopcroft} of reading in a word of finite
length and only at the end deciding to accept or reject. This aspect of our
model is a consequence of reading in process languages which are subword
closed.

In either the recognition or generation case, we will
discuss only models for arbitrarily long, but finite-time observations.
This circumvents several technical issues that arise with recognizing
and generating infinite-length strings, which is the subject of
$\omega$-language theory of B\"uchi automata \cite{Thom90a}.

Part of the burden of the following sections is to introduce a number
of specializations of stochastic machines. Although it is rarely good
practice to use terminology before it is defined, in the present setting
it will be helpful when tracking the various machine types to explain our
naming and abbreviation conventions now.

In the most general case---in particular, when the text says
nothing else---we will discuss, as we have just done, \emph{machines}.
These are input-output devices or transducers and we will denote this
in any abbreviation with a capital T. These will be specialized to
\emph{recognizers}, abbreviated R, and \emph{generators}, denoted G.
Within these basic machine types, there will be various alternative
implementations. We will discuss \emph{stochastic} (S) and \emph{quantum}
(Q) versions. Within these classes we will also distinguish the additional
property \emph{determinism}, denoted D. 

As noted above, the entire development concerns machines with a finite
set of states. And so, we will almost always drop the adjectives
``finite-state'' and ``finitary'', unless we wish to emphasize these
aspects in particular.

\subsection{Stochastic Recognizers}
\label{sec:StochasticRecognizers}

Stochastic devices that recognize inputs have been variously defined since the
first days of automata theory. Rabin's probabilistic automata \cite{rabi63},
for example, associate a stochastic matrix to each input symbol so that for a
given state and input symbol the machine stochastically transitions to a
successor state. Accepting an input string $x^L$ \emph{with cut point}
$\lambda$ is defined operationally by repeatedly reading in the same string
and determining that the acceptance probability was above threshold:
$p(x^L) > \lambda$. Accepting or rejecting \emph{with isolated cut point}
$\lambda$ is defined for some $\delta > 0$ with
$|p(x^L) - \lambda| \leq \delta$, respectively.

Here we introduce a \emph{stochastic recognizer} that applies a variable
form of cut-point recognition to process languages with the net effect
of representing the word distribution within a uniform tolerance.

One difference between the alternative forms of acceptance is the normalization
over equal-length strings for stochastic language recognition. Thus, Rabin's
probabilistic automata do not recognize stochastic languages, but merely
assign a number between $0$ and $1$ to each word being read in.
The same is true for Paz's stochastic sequential machines.

\begin{Def}
\label{def:sfr}
A \emph{stochastic finite-state recognizer} (SR) is a stochastic transducer
with $|Y| = 1$ and $T(y|x) = T(x)$.
\end{Def}

One can think of the output symbol as $accept$. If no symbol is output the
recognizer has halted and rejected the input.

An SR's state-to-state transition matrix:
\begin{align}
\label{eq:StateToStateSR}
T = \sum_{x\in X} T(x)~,
\end{align}
is a stochastic matrix.

\begin{Def}
An SR \emph{accepts} a process language $\prcl$ with threshold $\delta$, if
and only if for all $w \in \prcl$
\begin{align}
|\Prob(w) - \braket{\pi^0 |  T(w)  | \eta }| \leq \delta
\end{align}
and for all $w \notin \prcl$, $\braket{\pi^0 |  T(w)  | \eta } = 0$.
\end{Def}

The first criterion for accepting a process language is that all words in the
language lead the machine through a series of transitions with positive
probability and that words not in the language are assigned zero probability.
That is, it accepts the support of the language. The second criterion is that
the probability of accepting a word in the language is equal to the word's
probability within
a threshold $\delta$. Thus, an SR not only tests for membership in a
formal language, it also recognizes a \emph{function}: the probability
distribution of the language. For example, if $\delta=0$ the SR accepts
exactly a process language's word distribution. If $\delta>0$ it accepts
the probability distribution with some fuzziness, still rejecting all
of the language's probability-$0$ words. As mentioned before, recognition
happens at each
time step. This means that in practice the experimenter runs an ensemble
of SRs on the same input. The frequency of acceptance can then be compared
to the probability of the input string computed from the $T(x)$.  

\begin{Def}
The \emph{stationary state distribution} $\bra{\pi^s }$, which gives the
asymptotic state visitation probabilities, is determined by the left
eigenvector of $T(x)$:
\begin{equation}
\bra{\pi^s } = \bra{\pi^s }  T(x)~,
\label{eqn:pibar}
\end{equation}
normalized in probability: $\sum_{i=0}^{|S|-1} \pi^s_i = 1$.
\end{Def}

For a series $ x_0 x_1 \cdots x_{L-1}$ of input symbols the action of the
corresponding SR upon acceptance is a product of transition matrices:
\begin{equation*}
\label{eqn:Tvu}
T(x^L) = T(x_0) T(x_1)\cdots T(x_{L-1}) ~,
\end{equation*}
whose elements $T_{ij}(x^L)$ give the probability of making a transition
from state $i$ to $j$ and generating output $accept$ when reading input $x^L$.
If the SR starts in state distribution $\bra{\pi^0 }$, the state distribution
$\bra{\pi(x^L) }$ after accepting word $x^L$ is
\begin{equation}
\bra{\pi(x^L)} = \bra{ \pi^0} T(x^L)  ~.
\end{equation}
In this case, the probability of accepting $x^L$ is
\begin{equation}
\label{eq:prxL}
\Prob(x^L) = \braket{\pi^0 |  T(x^L)  | \eta }~. 
\end{equation}

We have the following special class of SRs.
\begin{Def}
\label{def:ssdfr}
A \emph{stochastic deterministic finite-state  recognizer} (SDR) is a
stochastic finite-state recognizer whose substochastic transition matrices
$T(x)$ have at most one nonzero element per row.
\end{Def}

A word accepted by an SDR is associated with one and only one path. This
allows us to give an efficient expression for the word distribution of
the language exactly ($\delta = 0$) recognized by an SDR:
\begin{equation}
\Prob(x^L) =
  T_{s_0 s_1} (x_0) T_{s_1 s_2} (x_1)
  \cdots
  T_{s_{L-1} s_{L}} (x_{L-1}) ~,
\label{eq:DetProbxL}
\end{equation}
where $s_1 s_2 \ldots s_L$ is the unique series of states along the
path selected by $x^L$.

There is an important difference here with Eq.~(\ref{eq:prxL}). 
Due to determinism, the
computational cost for computing the word probability $\Prob(x^L)$
from SDRs increases only linearly with $L$; whereas it is exponential
for SRs.

\begin{figure}  
\begin{center}
\resizebox{2.7in}{!}{\includegraphics{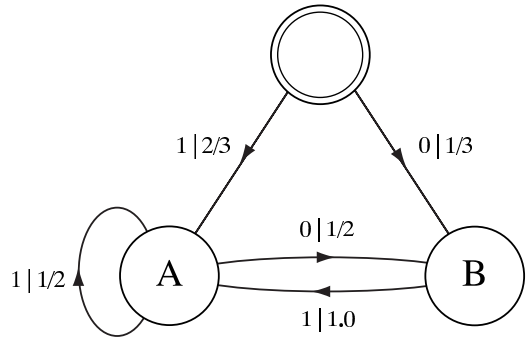}}
\end{center}
\caption{Stochastic deterministic recognizer for the Golden Mean
  process language of Fig.~\ref{fig:stol}. The edges are labeled $x|p$,
  where $x \in X$ and $p = T_{ij}(x)$. The start state $\bra{\pi^0} =
  (1,0,0)$ is double circled. The reject state and all transitions to it are
	omitted; as is the output $accept$ on all edges.
  }
\label{fig:sfr}
\end{figure}

Figure~\ref{fig:sfr} shows an example of an SDR that recognizes the Golden
Mean process language. That is, it rejects any word containing two consecutive
$0$s and accepts any other word with nonzero probability. This leads, in turn,
to the self-similar structure of the support of the word probability
distribution noted in Fig.~\ref{fig:stol}.

A useful way to characterize this property is to list a process language's
\emph{irreducible forbidden words}---the shortest disallowed words. In the
case of the Golden Mean formal language, this list has one member:
$\mathcal{F} = \{00\}$. Each irreducible word is associated with a family of
longer words containing it. This family of forbidden words forms a Cantor set
in the space of sequences, as described above. (Recall Fig.~\ref{fig:stol}.)

If we take the threshold to be $\delta = 0$, then the SDR recognizes only the
process language shown in Fig.~\ref{fig:stol}. If $\delta = 1$, in contrast,
the SDR would accept process languages with any distribution on the Golden Mean
process words. That is, it always recognizes the language's support.

One can easily calculate word probabilities and state distributions
for the Golden Mean Process using the SDR's matrix representation.  
\begin{equation}
\label{eqn:tgm}
T(0) = \left(\begin{array}{ccc} 
0 & 0 & \tfrac{1}{3} \\ 
0 & 0 & \tfrac{1}{2} \\ 
0 & 0 & 0
\end{array}\right) ~\mathrm{and}~
T(1) = \left(\begin{array}{ccc} 
0 & \tfrac{2}{3} & 0 \\ 
0 & \tfrac{1}{2} & 0 \\ 
0 & 1 & 0
\end{array}\right).
\end{equation}
We use Eq.~(\ref{eq:prxL}) with the start state distribution $\bra{\pi^0} =
(1,0,0)$ to calculate the $L = 1$ word probabilities:
\begin{align}
\nonumber
\Prob(0) & = \braket{\pi^0 |  T(0)  | \eta }
	= \tfrac{1}{3}~,\\
\nonumber
\Prob(1) & = \braket{ \pi^0 |  T(1)  | \eta }
	= \tfrac{2}{3}~.
\end{align}
(Eq.~(\ref{eq:DetProbxL}) would be equally applicable.)
At $L = 3$ one finds for $x^3 = 011$:
\begin{equation}
\nonumber
\Prob(011) = \braket{ \pi^0 |  T(011)  | \eta }
 = \braket{ \pi^0 |  T(0)  T(1)  T(1)  | \eta }
 = \tfrac{1}{6}.
\end{equation}
In fact, all $L=3$ words have the same probability, except for $x^3 = 101$,
which has a higher probability, $\Prob(101) = \tfrac{1}{3}$, and
$x^3 \in \{000, 001, 100\}$, for which $\Prob(x^3) = 0$. (Cf. the $L = 3$
word distribution in Fig.~\ref{fig:stol}.)

The conditional probability of a $1$ following a $0$, say, is calculated in
a similarly straightforward manner: 
\begin{equation}
\nonumber
\Prob(1|0) = \frac{\Prob(01)}{\Prob(0)}
	= \frac{\braket{ \pi^0 |  T(0)  T(1)  | \eta }}
	{ \braket{ \pi^0 |  T(0)  | \eta }}
	= 1~.
\end{equation}
Whereas, the probability $\Prob(0|0)$ of a $0$ following a $0$ is zero,
as expected.

\subsection{Stochastic Generators}
\label{sec:StochasticGenerators}

As noted in the introduction, finite-state machines generating strings
of symbols can serve as useful models for structure in dynamical systems.
They have been used as computational models of classical dynamical systems
for some time; see Refs.
\cite{Gras86,crutchfield:89,Fras90b,Auer90a,crutchfield:93,Yi94a,lind},
for example.

As we also noted, automata that only generate outputs are less often encountered
in formal language theory \cite{hopcroft} than automata operating as
recognizers. One reason is that redefining a
conventional recognizer to be a device that generates output words is
incomplete. A mechanism for choosing which of multiple transitions to
take when leaving a state needs to be specified. And this leads naturally
to probabilistic transition mechanisms, as one way of completing a
definition. We will develop finite-state generators by paralleling the
development of SRs.

\begin{Def}
\label{def:sfg}
A \emph{stochastic finite-state generator (SG)} is a stochastic transducer with $|X| = 1$.
\end{Def}
The input symbol can be considered a clock signal that drives the machine from
state to state. The transition matrices can be simplified to $T(y) = T(y|x)$.
An SG's state-to-state transition probabilities are given by the stochastic
\emph{state-to-state transition matrix}:
\begin{equation}
\label{eq:StateToStateSG}
T = \sum_{y\in Y} T(y)~. 
\end{equation}
Word probabilities are calculated as with SRs, save that one exchanges
input symbols $x$ with output symbols $y$:
\begin{equation}
\label{eq:pryl}
\Prob(y^L) = \braket{\pi^0 |  T(y^L)  | \eta }~. 
\end{equation}

We define the following special class of SGs.
\begin{Def}
A \emph{stochastic deterministic finite-state generator} (SDG) is a stochastic
finite-state generator in which each matrix $T(y)$ has at most one nonzero
entry per row.
\label{def:dG}
\end{Def}

As with SDRs, given the generator's state and an output symbol, the next state
is uniquely determined. Similarly, it is less costly to compute word
probabilities:
\begin{equation}
\Prob(y^L) =
  T_{s_0 s_1} (y_0) T_{s_1 s_2} (y_1)
  \cdots
  T_{s_{L-1} s_{L}} (y_{L-1}) ~.
\label{eq:DetProbyL}
\end{equation}
Given an initial state distribution, a sum is taken over states, weighted by
their probability. Even so, the computation increases only linearly with $L$.
In the following we concentrate on SDGs.

As an example, consider the generator for the Golden Mean process language.
Its matrix representation is the same as for the Golden Mean recognizer given
in Eqs.~(\ref{eqn:tgm}). Its graphical representation is the same as in
Fig.~\ref{fig:sfr}, except that the edge labels $x|p$ there should be given
as $p|y$. (We return to the relationship between recognizers and equivalent
generators shortly.) It turns out this is the smallest generator, but the
proof of this will be presented elsewhere.

One can easily calculate word probabilities and state distributions for the
Golden Mean Process using the SDG's matrices. Let us consider a method,
different from that used above for SRs, that computes probabilities using
the asymptotically
recurrent states only. This is done using the stationary state distribution
and the transition matrices restricted to the asymptotically recurrent states.
The method is useful whenever the start state is not known, but the asymptotic
behavior of the machine is. The transition matrices for the SDG, following
Eqs.~(\ref{eqn:tgm}), become:
\begin{equation}
T(0) = \left(\begin{array}{cc} 0 & \frac{1}{2} \\ 0 & 0
\end{array}\right) ~\mathrm{and}~
T(1) = \left(\begin{array}{cc} \frac{1}{2} & 0 \\ 1 & 0
\end{array}\right).
\end{equation}
The stationary state distribution $\bra{ \pi^s }$ is the
left eigenvector of the state-to-state transition matrix $T$,
Eq.~(\ref{eq:StateToStateSG}):
$ \bra{\pi^s} = \bra{\tfrac{2}{3},\tfrac{1}{3}}$.

Assuming that the initial state is not known, but the process has been
running for a long time, we use Eq.~(\ref{eq:pryl}) with $\bra{ \pi^s }$
to calculate the $L = 1$ word probabilities:
\begin{align}
\nonumber
\Prob(0) & = \braket{\pi^s |  T(0)  | \eta }
	= \tfrac{1}{3}~,\\
\nonumber
\Prob(1) & = \braket{ \pi^s |  T(1)  | \eta }
	= \tfrac{2}{3}~.
\end{align}
At $L = 3$ one finds for $y^3 = 011$:
\begin{equation}
\nonumber
\Prob(011) = \braket{ \pi^s |  T(011)  | \eta }
 = \braket{ \pi^s |  T(0)  T(1)  T(1)  | \eta }
 = \tfrac{1}{6} ~.
\end{equation}
All $L=3$ words have the same probability, except for
$y^3 = 101$, which has a higher probability, $\Prob(101) = \tfrac{1}{3}$,
and $y^3 \in \{000, 001, 100\}$, for which $\Prob(y^3) = 0$.
(Cf. the $L = 3$ distribution in Fig.~\ref{fig:stol}.)

These are the same results found for the Golden Mean Process recognizer.
There, however, we used a different initial distribution.
The general reason why these two calculations lead to the same result is not
obvious, but an explanation would take us too far afield.

\begin{figure}  
\begin{center}
\resizebox{2.7in}{!}{\includegraphics{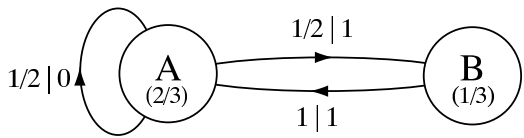}}
\end{center}
\caption{A deterministic generator of the Even Process: Blocks of an even
  number of $1$s are separated by $0$s. Only the asymptotically recurrent
  states are shown.
  Edges are labeled $p \,| \, y$, where $y \in Y$ and $p = T_{ij}(y)$. 
  The numbers in parentheses give a state's asymptotic probability.
  }
\label{fig:even}
\end{figure}

As a second example of an SDG consider the Even Process whose
language consists of blocks of even numbers of $1$s bounded by $0$s.
The substochastic transition matrices for its recurrent states are
\begin{align}
\label{eqn:t11}
T(0) = \left(\begin{array}{cc} \frac{1}{2} & 0 \\ 0 & 0
  \end{array}\right) ~\mathrm{and}~
T(1) = \left(\begin{array}{ccc} 0 & \frac{1}{2} \\ 1 & 0
  \end{array}\right) ~.
\end{align}
The corresponding graph is shown in Fig.~\ref{fig:even}. Notice that the
state-to-state transition matrix $T$ is the same as the previous model
of the Golden Mean Process. However, the Even Process is substantially
different; and its SDG representation lets us see how. 
The set of irreducible forbidden words is countably infinite \cite{weis73}:
$\mathcal{F} = \{ 01^{2k+1}0: ~k = 0, 1, 2, \ldots \}$. Recall that the Golden
Mean Process had only a single irreducible forbidden word $\{00\}$. One
consequence is that the words in the Even Process have a kind of infinite
correlation: the ``evenness'' of the length of $1$-blocks is respected
over arbitrarily long words. This makes the Even Process effectively
non-finite: As long as a sequence of $1$s is produced, memory of the
initial state distribution persists. Another difference is that the support
of the word distribution has a countable infinity of distinct Cantor
sets---one for each irreducible forbidden word.  Thus, the Even Process falls
into the broader class of \emph{finitary} processes.

\subsection{Properties}
\label{sec:SRSGProp}

We can now describe the similarities and differences between stochastic
and other kinds of recognizers and between the various classes of generators.
Let $\stochL(\mach)$ denote the stochastic language recognized or generated by
automaton $\mach$. Let $\PRCL(C)$ denote the \emph{set} of stochastic languages
generated or recognized by machines in class $C$.

The relationships between the languages associated with the various machine
types follow rather directly from their definitions. We swap input and output
alphabets and reinterpret the same transition matrices, either as specifying
$x|p$ or $p|y$ as required. All, that is, except for the last two results,
which may be unexpected.

\begin{Prop}
\label{prop:rrl}
For every SR, $\supp \stochL(SR)$ is a regular language.
\end{Prop}

\begin{ProProp}
The graph of an SR, removing the probabilities, defines a finite-state
recognizer and accepts, by definition, a regular language \cite{hopcroft}.
This regular language is the support of $\stochL(SR)$ by construction.
\end{ProProp}

\begin{Prop}
\label{prop:rprcl}
For every SR, $\stochL(SR)$ is a process language.
\end{Prop}

\begin{ProProp}
The first property to establish is that the set of words recognized by an SR is
subword closed: if $\Prob(x^L) > 0$, then all $w \in \mathrm{sub}(x^L)$ have
$\Prob(w) > 0$. This is guaranteed by definition, since the first input symbol
not encountering an allowed transition leads to rejection of the whole input,
see the SR definition.

The second property to establish is that the word distribution $\Prob(x^L)$
is normalized at each $L$. This follows from $T$ in
Eq.~\ref{eq:StateToStateSR} being stochastic.
\end{ProProp}

\begin{Prop}
\label{prop:SRSG}
SGs and SRs generate and recognize, respectively, the same set of languages:
$\PRCL(SG) = \PRCL(SR)$. 
\end{Prop}

\begin{ProProp}
Consider SG's transition matrices $T(y)$ and form a new set $T(x)$
in which $X = Y$. The $T(x)$ define an SR that recognizes $\stochL(SG)$.

It follows that $\PRCL(SG) \subseteq \PRCL(SR)$.

Now consider SR's transition matrices $T(x)$ and form a new set $T(y)$
in which $Y = X$. The $T(y)$ define an SG that generates $\stochL(SR)$.

It follows that $\PRCL(SG) = \PRCL(SR)$.
\end{ProProp}

\begin{Cor}
\label{cor:SGreg}
For every SG, $\supp \stochL(SG)$ is a regular language.
\end{Cor}

\begin{Cor}
\label{cor:SGproc}
For every SG, $\stochL(SG)$ is a process language.
\end{Cor}

\begin{Cor}
\label{cor:SDRSDG}
SDGs and SDRs generate and recognize, respectively, the same set of languages:
$\PRCL(SDG) = \PRCL(SDR)$. 
\end{Cor}

These equivalences are intuitive and expected. They do not, however,
hint at the following, which turn on the interplay between nondeterminism
and stochasticity.

\begin{Prop}
There exists an SG such that $\prcl(SG)$ is not recognized by any SDR.
\label{prop:SDRsubsetG}
\end{Prop}

\begin{figure}  
\begin{center}
\resizebox{3.0in}{!}{\includegraphics{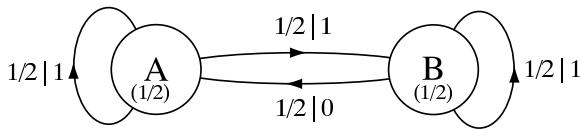}}
\end{center}
\caption{A nondeterministic generator that produces a process language not
  recognized by any (finite-state) SDR. Only asymptotically recurrent states
are shown.
  Edges are labeled $p \,| \, y$, where $y \in \{0,1\}$ and $p = T_{ij}(y)$. 
  }
\label{fig:g_sns}
\end{figure}

\begin{ProProp}
We establish this by example.
Consider the nondeterministic generator in Fig.~\ref{fig:g_sns}, the
\emph{Simple Nondeterministic Source} (SNS). 
To show that there is no possible construction of an SDR we
argue as follows. If a $0$ appears, then the
generator is in state $\mathrm{A}$. Imagine this is then followed by
a block $1^k$. At each $k$ the generator is in either state $\mathrm{A}$
or $\mathrm{B}$. The probability of seeing a $0$ next is ambiguous
(either $0$ or $1/2$) and depends on the exact history of internal states
visited. Deterministic recognition requires that a recognizer be in a
state in which the probability of the next symbol is uniquely given. While
reading in $1$s the recognizer would need a new state for each $1$ connecting
to the same state (state $\mathrm{A}$) on a $0$. 
Since this is true for all $k$, there is no finite-state SDR that
recognizes the SNS's process language.
\end{ProProp}

Ref. \cite{crut94} gives an SDR for this process that
is minimal, but has a countably infinite number of states. Note that
$\supp{\prcl(SNS)}$ is the support of the Golden Mean process language.

\begin{Cor}
There exists an SR such that $\prcl(SR)$ is not generated by any SDG.
\label{DGsubsetSR}
\end{Cor}

These propositions say, in essence, that deterministic machines generate or
recognize only a subset of the finitary process languages. In particular, Props.
\ref{prop:SRSG}, \ref{prop:SDRsubsetG}, and Cor.~\ref{cor:SDRSDG} imply proper
containment: $\PRCL(SDR) \subset \PRCL(SG)$ and $\PRCL(SDG) \subset \PRCL(SR)$.
This is in sharp contrast with the standard result in formal language theory:
deterministic and nondeterministic automata recognize the same class of
languages---the regular languages \cite{hopcroft}.

This ends our development of classical machines and their specializations.
We move on to their quantum analogs, following a strategy that is familiar
by now.



\section{Finitary Quantum Processes}
\label{sec:fqp}

As with stochastic processes, the evolution of a quantum system is monitored
by a series of measurement outcomes---numbers registered in some way. Each
outcome can be taken to be the realization of a random variable. The
distribution over sequences of these random variables is what we call a
\emph{quantum process}. We will consider the finitary version of quantum
processes in the same sense as used for the classical stochastic processes:
The internal resources used during the evolution are finitely specified.


\subsection{Quantum States}
\label{sec:siqs}

Quantum mechanics is sometimes viewed as a generalization of classical
probability theory with noncommuting probabilities. It is helpful, therefore,
to compare classical stochastic automata and quantum automata and, in
particular, to contrast the corresponding notions of state. The goal is to
appreciate what is novel in quantum automata. The reader should have a working
knowledge of quantum mechanics at the level of, say, Ref. \cite{shankar}.

In the classical (stochastic) setting an automaton has internal
states $S$ and also a distribution $\bra{\pi}$ over them. The distribution
itself can be taken to be a ``state'', but of what? One interpretation
comes from considering how an observer monitors a series of
outputs from a stochastic generator and predicts, with each observed symbol,
the internal state $s \in S$ the automaton is in. This prediction is a
distribution $\bra{\pi}$ over the internal states---one that represents the
observer's best guess of the automaton's current internal state. In this
sense the distribution is the state of the best predictor. If
$\bra{\pi} = (0, \ldots, 0, \pi_i = 1, 0, \ldots, 0)$, then the observer
knows exactly what internal state, $s_i \in S$, the automaton is in. For these
special cases one can identify state distributions and internal states.

Similarly, there are several kinds of state that one might define for a
quantum automaton. Each quantum automaton will consist of \emph{internal
states} and we will take the \emph{state of the automaton} to be a
\emph{superposition} over them. The central difference with
classical (stochastic) automata is that the superposition over internal
states is not a probability distribution. In particular, internal states
have complex amplitudes and, therefore, they potentially
interfere. This, in turn, affects the process language associated with the
quantum automaton.

In contrast with quantum automata, the state of a quantum dynamical system
depends on the choice of a basis that spans its state space. The state is
completely specified by the system's \emph{state vector}, a unit vector
represented as a sum of \emph{basis states} that span the state space.
However, if one chooses a basis consisting of the eigenstates of an observable
and associates them with internal states of quantum automaton, there is a
simple correspondence between a state vector of a quantum dynamical system
(a superposition of basis states) and a state of a quantum automaton (a
superposition over internal states). Thus, we will use the terms
\emph{internal states} (of an automaton) and \emph{basis states} (of a
quantum dynamical system's state space) interchangeably. By similar reasoning,
the \emph{state vector} (of a quantum dynamical system) and \emph{state}
(of a quantum automaton) will be used interchangeably.

In the vocabulary of quantum mechanics, at any moment in time a given quantum
automaton is in a \emph{pure state}---another label for a superposition over
internal states. An observer's best guess as to the automaton's current pure
state is a probability distribution over state vectors---the so-called
\emph{mixed state}. 

It is helpful to imagine a collection of individual quantum automata,
each in a (pure) state, that is specified by a distribution of weights.
One can also imagine a single quantum automaton being in different pure
states at different moments in time. The time-averaged state then is also
a mixed state. It is the latter picture that we adopt here.

The fact that a quantum pure state can be a superposition of basis states is
regarded as the extra structure of quantum mechanics that classical mechanics
does not have. We respect this distinction by building a
hierarchy of quantum states that goes from basis states to superpositions of
basis states to mixtures of superpositions. The analogous classical-machine
hierarchy goes only from internal states to distributions over internal states.

\subsection{Quantum Measurement}
\label{sec:meas}

We now turn to the measurement process, a crucial and also distinctive
component in the
evolution of a quantum dynamical system, and draw parallels with quantum
automata. In setting up an experiment, one makes choices of how and when to
measure the state of a quantum system. These choices typically affect what one
observes, and in ways that differ radically from classical dynamical systems.

Measurement is the experimental means of characterizing a system in the sense
that the observed symbols determine the process language and any subsequent
prediction of the system's behavior. The measurement of a quantum mechanical
system is described by a Hermitian operator that projects the
current state onto one (or several) of the operator's eigenstates. After a
measurement, the system is, with certainty, in the associated (subset of)
eigenstate(s). Such an operator is also called an \emph{observable} and the
\emph{eigenvalues} corresponding to the eigenstates are the observed
measurement \emph{outcomes}.

To model this situation with a quantum automaton, we identify the states of
the automaton with the eigenstates of a particular observable. A measurement
is defined through an operator that projects the automaton's current state
vector onto one (or a subset) of its internal (basis) states. The ``observed''
measurement outcome is emitted as a symbol labeling the transition(s) which
enter that internal state (or that subset of states). 


\section{Quantum Transducers}

The study of quantum finite-state automata has produced a veritable zoo of
alternative models for language recognition. (These are reviewed below in
Section~\ref{sec:altqr}.) Since we are interested in recognition, generation,
and transduction of process languages, we start out defining a generalized
quantum-finite state transducer and then specialize. We develop a series of
quantum finite-state automaton models that are useful for recognition and
generation and, ultimately, for modeling intrinsic computation in finitary
quantum processes. It is worth recalling that these quantum finite-state
machines form the lowest level of a hierarchy of quantum computational models.
Thus, they are less powerful than quantum Turing machines. Nevertheless, as
we will see, they exhibit a diversity of interesting behaviors. And, in
any case, they represent currently feasible quantum computers.

\subsection{Definition}

We define a quantum transducer that corresponds to the standard
quantum mechanical description of a physical experiment.
\begin{Def}
\label{def:qa}
A \emph{quantum transducer} (QT) is a tuple
$\{Q,\bra{\psi} \in \mathcal{H},X,Y,\mathbf{T}(Y|X) \}$ where
\begin{enumerate}
\setlength{\topsep}{0pt}
\setlength{\itemsep}{0pt}
\item $Q = \{q_i: i = 0, \ldots, n-1 \}$ is a set of $n$ \emph{internal states}.
\item The \emph{state vector} $\bra{\psi}$ lies in an $n$-dimensional Hilbert
	space $\mathcal{H}$; its initial value is the \emph{start state}
	$\bra{\psi^0}$.
\item $X$ and $Y$ are finite alphabets for input and output symbols,
	respectively.
\item $\mathbf{T}(Y|X)$ is a set of $n\times n$ \emph{transition matrices}
	$\{T(y|x) = U(x) P(y), x \in X, y \in Y \}$ that are products of 
\begin{enumerate}
	\setlength{\topsep}{0pt}
	\setlength{\itemsep}{0pt}
  \item a unitary matrix $U(x)$: $U^\dagger (x) = U^{-1} (x)$
	($\dagger$ denotes complex transpose);
	and
  \item a projection operator $P(y)$.
\end{enumerate}
\end{enumerate}
At each time step a quantum transducer ($QT$) reads a symbol $x \in X$ from the input, outputs a
symbol $y \in Y$, and updates its state vector $\bra{\psi}$ via $T(y|x)$.
\end{Def}

The preceding discussion of state leads to the following correspondence
between a QT's internal states and state vectors.
\begin{Def}
\label{def:basisstates}
One associates an \emph{internal state} $q_i \in Q$ with the eigenstate
$\bra{\phi_i}$ of an observable such that:
\begin{enumerate}
\setlength{\topsep}{0pt}
\setlength{\itemsep}{0pt}
\item 
	For each $q_i \in Q$ there is a basis vector
	$\bra{\phi_i} = (0, \ldots, 1, \ldots, 0 )$
	with a $1$ in the $i^\mathrm{th}$ component.
\item The set $\{\bra{ \phi_i} : i=0,1,\dots,n-1 \}$ spans the Hilbert space
	$\mathcal{H}$.
\end{enumerate}
\end{Def}

\begin{Def}
\label{def:statevector}
A \emph{state vector} $\bra{\psi} \in \mathcal{H}$ is a unit vector.
It can be expanded in terms of basis states $\bra{\phi_i}$:
\begin{equation}
	\bra{\psi}= \sum_{i=0}^{n-1} \bra{\phi_i} c_i ~,
\end{equation}
with $c_i \in \field{C}$ and $\sum_{i=0}^{n-1} c_i^*c_i = 1$.
\end{Def}

Identifying internal states $q_i$ and basis states $\bra{\phi_i}$ connects the
machine view of a quantum dynamical system with that familiar from standard
developments of quantum mechanics. A QT state is given by its current state
vector $\bra{\psi}$. At each time step a symbol $x$ is read in, which selects
a unitary operator $U(x)$. The operator is applied to the state vector and the
result is measured via $P(y)$. The output, an eigenvalue of the observable,
is symbol $y$.

We describe a QT's operation via the evolution of a bra (row) vector. We make
this notational choice, which is unconventional in quantum mechanics, for two
reasons that facilitate comparing classical and quantum automata. First, the
state distribution of a classical finite-state machine is given conventionally
by a row vector. And second, the graphical meaning of a transition from state
$i$ to $j$ is reflected in the transition matrix entries $T_{ij}$, only if
one uses row vectors and left multiplication with $T$. This is also
convention for stochastic processes.

\subsection{Measurement}

The projection operators are familiar from quantum mechanics and can be
defined in terms of the internal (basis) states as follows.
\begin{Def}
A \emph{projection operator} $P(y)$ is the linear operator
\begin{equation}
P(y) = \ket{\phi_{i}}\bra{\phi_{i}}~,
\end{equation}
where $\phi_i$ is the eigenvector of the observable with eigenvalue $y$.
In the case of degeneracy $P(y)$ sums over a complete set $\{ i \}$ of
mutually orthogonal eigenstates: 
\begin{equation}
P(y) = \sum_{ \{ i \} } \ket{\phi_{i}} \bra{\phi_{i}} ~.
\end{equation}
Each $P$ is Hermitian ($P^{\dagger} = P$) and idempotent ($P^2 = P$).
\end{Def}

$\mathbf{P} \equiv \{ P(y) : y \in Y\cup \{\lambda\} \}$ is the set of
projection operators with $\sum_{y\in Y} P(y) = {\mathbbm 1}$,
where ${\mathbbm 1}$ is the identity matrix. $\lambda$ is the null symbol and
a placeholder for ``no measurement''. We take $P(\lambda) = {\mathbbm 1}$ and
do not include it in the calculation of word probabilities, for example.
``No measurement'' differs from a \emph{non-selective} measurement where a
projection takes place, but the outcome is not detected. The decision whether
to perform a measurement or not is considered an input to the QT. 

In the eigenbasis of a particular observable the corresponding matrices only
have $0$ and $1$ entries. In the following we assume such a basis. In addition,
we consider only projective measurements which apply to closed quantum
systems. (Open systems will be considered elsewhere.)

In quantum mechanics, one distinguishes between \emph{degenerate} and
\emph{non-degenerate} measurement operators \cite{cohen}. A
\emph{non-degenerate measurement operator} projects onto one-dimensional
subspaces of $\mathcal{H}$. That is, the eigenvectors of the operator  all have
distinct eigenvalues. In contrast, the operators associated with a
\emph{degenerate measurement} have degenerate eigenvalues. Such an operator
projects onto higher-dimensional subspaces of $\mathcal{H}$. After such a
measurement the QT is potentially in a 
superposition of states $\sum_{i} c_{i}\bra{\phi_{i}}$, where $i$ 
sums over the degenerate set of mutually orthogonal eigenstates.
Just as degeneracy leads to interesting
consequences in quantum physics, we will see in the examples to follow that
degenerate eigenvalues lead to interesting quantum languages.

QTs model a general experiment on a quantum dynamical system. As such they
should be contrasted with the sequential machines and transducers of
Refs.~\cite{gudd99} and \cite{freivalds:01}, respectively, that map the
current quantum state onto an output. This mapping, however, is not associated
with a measurement interaction and lacks physical interpretation.

\subsection{Evolution and Word Distributions}

We can now describe a QT's operation as it scans its input. Starting in
state $\bra{\psi^0}$ it reads in a symbol $x \in X$ from an input word
and updates its state by applying the unitary matrix $U(x)$. Then the state
vector is projected with $P(y)$ and renormalized. Finally, symbol $y \in Y$
is emitted. That is, the state vector after a single time-step of a QT is
given by:
\begin{align}
\label{eq:psiy|x}
\notag
\bra{\psi(y|x)}  & = 
  \frac{\bra{\psi^0} T(y|x)}
  {\sqrt{\braket{\psi^0\vert T(y|x) T^{\dagger}(y|x)\vert\psi^0}}} \\
  & = 
  \frac{\bra{\psi^0} U(x)  P(y)  }
  {\sqrt{\braket{ \psi^0\vert U(x)P(y) U^{\dagger}(x) \vert\psi^0}}}
  ~.
\end{align}
In the following we drop the renormalization factor in the denominator to
enhance readability. It will be mentioned explicitly when a state is not to
be normalized.

When a QT reads in a length-$L$ word $x^L \in X^L$ and outputs a
length-$L$ word $y^L \in Y^L$, the transition matrix becomes
\begin{equation}
T(y^L|x^L) = U(x_0) P(y_0) \cdots U(x_{L-1}) P(y_{L-1})
\end{equation}
and the updated state vector is
\begin{equation}
\label{eq:psiyL|xL}
\bra{\psi(y^L|x^L)} = \bra{\psi^0} T(y^L|x^L)  ~. 
\end{equation}

Starting the QT in $\bra{\psi^0}$ the conditional probability $\Prob(y|x)$ of
the output symbol $y$ given the input symbol $x$ is calculated from the
state vector in Eq.~(\ref{eq:psiy|x}), before renormalization:
\begin{equation}
\label{eqn:pryx}
\Prob(y|x) =   \braket{{\psi(y|x)} | {\psi(y|x)}}~.
\end{equation}
The probability $\Prob(y^L|x^L)$ of output sequence $y^L$ conditioned on input
sequence $x^L$ is calculated similarly using Eq.~(\ref{eq:psiyL|xL}):
\begin{equation}
\label{eqn:pryLxL}
\Prob(y^L|x^L) =  \braket{{\psi(y^L|x^L)} | {\psi(y^L|x^L)}}~.
\end{equation}

\subsection{Properties}
\label{sec:STProp}

We draw out several properties of QTs on our way to understanding their
behavior and limitations.
\begin{Prop}
\label{cor:maxA}
A QT's output alphabet size is bounded: $|Y| \leq \dim (\mathcal{H})$.
\end{Prop}
\begin{ProProp}
This follows from the QT definition since output symbols
are directly associated with eigenvalues. The number of eigenvalues is
bounded by the dimension of the Hilbert space.
\end{ProProp}

Many properties of QTs are related to a subclass of STs, those with doubly
stochastic transition matrices. Given this, it is useful to recall the
relationship between unitary and doubly stochastic matrices.

\begin{Def}
Given a unitary matrix $U$, matrix $M$ with $M_{ij} = |U_{ij}|^2$ is
called a \emph{unistochastic} matrix.
\label{def:ustoc}
\end{Def}

A unistochastic matrix is doubly stochastic, which follows from the properties
of unitary matrices. Compared to a stochastic transducer, a QT's structure is
constrained through unitarity and this is reflected in its architecture. A
\emph{path} exists between node $i$ and node $j$ when $M_{ij}>0$. An equivalent
description of a quantum transducer is given by its graphical representation.

Recalling the types of graph state defined in Sec.~\ref{sec:graph}, we find
that only a subset occur in QTs. Specifically, a QT has no transient states.
\begin{Prop}
\label{prop:conn}
Every node $i$ of $\Graph$(QT), if connected to a set of nodes $j \neq i$, is
a member of a strongly connected set.
\end{Prop}

\begin{ProProp}
Given that one path exists from (say) $i$ to $j$, we must show that the
reverse one exists, going from $j$ to $i$. According to the definition of
a path it is sufficient to show this for the unistochastic matrix
$M_{ij} = |U_{ij}|^2$. A doubly
stochastic matrix can always be expressed as a linear combination of permutation
matrices. Thus, any vector $(0,0,\dots ,1, \dots)$ with only one $1$ entry can
be permuted into any other vector with only one $1$ entry. This is equivalent
to saying that, if there is a path from node $i$ to $j$ there
is a path from $j$ to $i$.
\end{ProProp}

The graph properties of a unitary matrix mentioned here should be compared
with those discussed by Severini \cite{seve03} and others. The graph of a
finite-state machine specified by a unitary matrix is a directed graph, or
\emph{digraph}. A digraph vertex is a \emph{source} (\emph{sink}) if it has
no ingoing (no outgoing) arcs. A digraph vertex is \emph{isolated} if it is
not joined to another. Ref.~\cite{seve03} characterizes these machines by
assuming their digraphs have no isolated nodes, no sinks, and no sources.
Given the preceding proposition the nonexistence of sinks or sources follows
simply from assuming no isolated nodes.

One concludes that QT graphs are a limited subset of digraphs, namely the
strongly connected ones. Furthermore, there is a constraint on incoming edges
to a node.
\begin{Prop}
\label{cor:edg}
All incoming transitions to an internal state are labeled with the same
output symbol.
\end{Prop}
\begin{ProProp}
Incoming transitions to internal state $q_i$ are labeled with output symbol
$y$ if $\bra{\phi_i}$ has eigenvalue $y$. Every eigenstate has a unique
eigenvalue, and so the incoming transitions to any particular state $q_i$ are
labeled with the same output symbol representing one eigenvalue. 
\end{ProProp}

\begin{Prop}
A QT's transition matrices $T(y|x)$ uniquely determine the unitary
matrices $U(x)$ and the projection operators $P(y)$.
\end{Prop}

\begin{ProProp}
Summing the $T(y|x)$ over all $y$ for each $x$ yields the unitary matrices
$U(x)$:
\begin{equation}
\sum_{y\in Y} T(y|x) = \sum_{y\in Y} U(x) P(y) = U(x)~.
\end{equation}
The $P(y)$ are obtained, from any of the $U(x)$, through the inverse of
$U^{-1}(x) = U^{\dagger}(x)$:
\begin{equation}
P(y) =   U^{\dagger}(x) T(y|x) ~.
\end{equation}

\end{ProProp}

\begin{Def}
A QT is \emph{reversible} if the automaton defined by the transpose of each
$U(x)$ and $P(y)$ is also a QT.
\end{Def}

\begin{Prop}
QTs are reversible.
\end{Prop}

\begin{ProProp}
The transpose of a unitary matrix is unitary. The transpose of a projection
operator is the operator itself.
\end{ProProp}

Graphically, the reversed QT is obtained by simply switching the direction of
the edges. This produces a transducer with the transition amplitudes $T_{ji}$,
formerly $T_{ij}$. The original input and output symbols,  which labeled
ingoing edges to state $q_i$, remain unchanged. Therefore, in general, the
languages generated by a QT and its reverse are not the same. By way of
contrast, this simple operation applied to an ST does not, in general, yield
another ST. A simple way to summarize these properties is that a QT forms a
group, an ST forms a semi-group.

\section{Quantum Recognizers and Generators}

The quantum transducer is our most general construct, describing a quantum
dynamical process in terms of inputs and outputs. We will now specialize
quantum transducers into recognizers and generators. We do this by paralleling
the strategy adopted for developing classes of stochastic transducers. For
each machine class we first give a general definition and then specialize, 
for example, yielding deterministic variants. We establish a number of
properties for each type and then compare their descriptive powers in terms
of the process languages each class can recognize or generate. The results are
collected together in a computational hierarchy of finitary stochastic and
quantum processes.

\subsection{Quantum Recognizers}
\label{sec:qfr}

Quantum finite-state machines are almost exclusively discussed as recognizing
devices. Following our development of a consistent set of quantum finite-state
transducers, we can now introduce quantum
finite-state recognizers as restrictions of QTs and compare these with
alternative models of quantum recognizers. Since we are interested in the
recognition of process languages our definition of quantum recognizers differs
from those introduced elsewhere; see Sec.~\ref{sec:altqr} below. The main
difference is the recognition of a process language including its word
distribution. The restrictions that will be imposed on a QT to achieve this are
similar to those of the stochastic recognizer.

\begin{Def}
\label{def:qfr}
A \emph{quantum finite-state recognizer} (QR) is a quantum transducer with
$|Y| = 1$ and $T(y|x) = U P(x) \equiv T(x)$.
\end{Def}

One can think of the output symbol $y$ as $accept$. The condition for
accepting a symbol is, then,
\begin{equation}
\Prob(x) = \braket{\psi^0 | T(x) T^{\dagger}(x)|\psi^0 }~.
\end{equation}

If no symbol is output the recognizer has halted and rejected the input.
Operationally, recognition works as it does in the classical setting. An
experimenter runs an ensemble of QRs on the same input. The frequency of
acceptance can then be compared to the probability of the input string
computed using the $T(x)$. 

\begin{Def}
A QR \emph{accepts} a process language $\prcl$ with word-probability threshold
$\delta$, if and only if for all $w\in \prcl$
\begin{equation}
\label{eq:qr}
  \left| \Prob(w) \! - \! \braket{\psi^0 | T(w) T^{\dagger}(w)|\psi^0 }\right|
  \leq \delta
\end{equation}
and for all $w \notin \prcl$, 
$\braket{\psi^0 | T(w) T^{\dagger}(w)|\psi^0 } = 0$.
\end{Def}

Acceptance or rejection happens at each time step. 

We also have deterministic versions of QRs.

\begin{Def}
\label{def:qdr}
A \emph{quantum deterministic finite-state recognizer} (QDR) is a quantum
recognizer with transition matrices $T(x)$ that
have at most one nonzero element per row.
\end{Def}

\subsection{Alternatives}
\label{sec:altqr}

Quantum finite automata were introduced by several authors in different ways,
and they recognize different classes of languages. To our knowledge the
first mention of \emph{quantum automata} was made by Albert in 1983
\cite{albe83}. Albert's results have been subsequently criticized by Peres as
being based on an inadequate notion of measurement \cite{pere84}. 

Kondacs and Watrous introduced 1-way and 2-way quantum finite-state automata
\cite{kondacs:97}. The $1$-way automata read symbols once and from left to
right (say) in the input word. Their 2-way automata scan the input word many
times moving either left to right or right to left. The automata allow for
measurements at every time step, checking for acceptance, rejection, or
continuation. They show that a 2-way QFA can recognize all regular languages
and some nonregular languages. 1-way
QFA are less powerful: They can only recognize a subset of the regular
languages. A more powerful generalization of a 1-way QFA is a 1-way QFA that
allows mixed states, introduced by Aharonov et al \cite{ahar98}. They also
allow for nonunitary evolution. Introducing the concept of mixed states simply
adds classical probabilities to quantum probabilities and is inherent in our
model of QTs. 

The distinctions between these results and the QRs introduced here largely
follow from the difference between regular languages and process languages.
Thus, the result in Ref.~\cite{kondacs:97} that no $1$-way quantum automaton
can recognize the language $(0+1)^*0$, does not apply to QTs. It clearly is
a regular language, but not a process language. Also, the result by Bertoni
and Carpentieri that quantum automata can recognize nonregular languages,
does not apply here \cite{bert01}. They find that a quantum automaton that is
measured only after the whole input has been read in can recognize a
nonregular language. A QR, however, applies measurement operators for every
symbol that is being read in.

Moore and one of the authors introduced 1-way quantum
automata (without using the term ``1-way'') \cite{moore:00}. It is less
powerful than the 1-way automata of Kondacs and Watrous, since it allows
only for a single measurement after the input has been read in. They also
introduced a generalized quantum finite-state automaton whose transition
matrices need not be unitary, in which case all regular languages are
recognized. A type of quantum transducer mentioned earlier, a
\emph{quantum sequential machine} was introduced by Gudder \cite{gudd99}.
The link, however, between machine output and quantum physical measurement
is missing. Freivalds and Winter introduced quantum transducers
\cite{freivalds:01} that at each step perform a measurement to determine
acceptance, rejection, or continuation of the computation. In addition,
they map the current quantum state onto an output. Here too, the mapping
is not associated with a measurement interaction and lacks physical
interpretation. 

These alternative models for quantum automata appear to be the most widely
discussed. There are others, however, and so the above list is by no means
complete. Our motivation to add yet another model of quantum finite-state
transducer and recognizer to this list is the inability of the alternatives
to recognize or process languages that represent quantum dynamical
systems subject to repeated measurement.

\subsection{Quantum Generators}
\label{sec:qg}

We now introduce quantum finite-state generators as restrictions of QTs and
as a complement to recognizers. They serve as a representation for the
behavior of autonomous quantum dynamical systems. In contrast to quantum
finite-state recognizers, quantum finite-state generators appear to not
have been discussed before. A quantum generator is a QT with only one
input. As in the classical case, one can think of the input as a clock signal
that drives the machine through its transitions. 

\begin{Def}
\label{def:qfg}
A \emph{quantum finite-state generator} (QG) is a quantum transducer with $|X|
= 1$.
\end{Def}

At each step it makes a transition from one state to another and emits a
symbol. As in the classical case there are nondeterministic (just implicitly
defined) and deterministic QGs. 
\begin{Def}
\label{def:qdg}
A \emph{quantum deterministic finite-state generator} (QDG) is a quantum
generator in which each matrix $T(y)$ has at most one nonzero entry per row.
\end{Def}

Interestingly, there is a mapping from a given QDG to a classical automaton.

\begin{Def}
\label{def:qecfg}
Given a QDG $\mach = \{U,P(y)\}$, the \emph{equivalent} (classical) SDG
$\mach^{\prime} = \{T(y)\}$ has unistochastic state-to-state transition
matrix $T$ with components $T_{ij} = [U_{ij}]^2$.
\end{Def}

We leave the technical interpretation of ``equivalent'' to
Thm.~\ref{the:qeDG} below.

As mentioned earlier, in quantum mechanics one distinguishes between degenerate
and non-degenerate measurements. Having introduced the different types of
quantum generators, we can now make a connection to degenerate measurements.

\begin{Def}
\label{def:cqg}
A \emph{quantum complete finite-state generator} (QCG) is a quantum generator
observed via non-degenerate measurements.
\end{Def}

In order to average over observations, we must extend the formalism of
quantum automata to describe distributions over state vectors. Recalling
the notions of state discussed in Section~\ref{sec:siqs}, this means we
need to describe mixed states and their evolution.

Let a system be described by a state vector $\bra{\psi_i}$ at time $t$. If we
do not know the exact form of $\bra{\psi_i}$ but only a set of possible
$\bra{\psi_i}, ~i = 0, ...,k-1$, then we give the best guess as to the system's
state in terms of a statistical mixture of the $\bra{\psi_i}$. This statistical
mixture is represented by a \emph{density operator} $\rho$ with weights $p_i$
assigned to the $\bra{\psi_i}$:
\begin{equation}
\rho = \sum_{i=0}^{k-1} p_i \ket{\psi_i}\bra{\psi_i} ~.
\end{equation} 
The main difference from common usage of ``mixed state'' is that we compare
the same state \emph{over time}; whereas, typically different \emph{systems}
are compared at a single time. Nevertheless, in both cases, the density matrix
formalism applies.

\subsection{Properties}

With this notation in hand, we can now establish a number of properties of
quantum machines.

\begin{Def}
A QG's \emph{stationary state} $\rho^s$ is the mixed state that is
invariant under unitary evolution and measurement:
\begin{equation}
 \rho^s = \sum_{y \in Y} P(y) U^{\dagger} \rho^s U P(y)~.
\end{equation}
\end{Def}
$\rho^s$ is the mixed state which the quantum machine is in \emph{on average},
since we are describing a single system that is always in a pure state. The
stationary state is therefore the best guess of an observer ignorant of the
machine's state. 

\begin{The}
\label{the:ss}
A QG's stationary state is the maximally mixed state:
\begin{equation}
 \rho^s =  n^{-1} \sum_{i=0}^{n-1} \ket{\phi_i}\bra{\phi_i} = \mathbbm{1}/n~.
\label{eq:rhos}
\end{equation}
\end{The}

\begin{ProThe}
Since the $\bra{\phi_i}$ are basis states, $\rho^s$ is a diagonal matrix equal
to the identity multiplied by a factor. Recall that the stationary distribution
of a Markov chain with doubly stochastic transition matrix is always uniform
\cite{cover}. And so, we have to establish that $\rho^s$ is an
invariant distribution:
\begin{align}
\rho^s & =  \sum_{y \in Y} P(y) U^{\dagger} \rho^s U P(y) \\
	& = n^{-1} \sum_{y \in Y} P(y) U^{\dagger} U P(y)\\
	& = n^{-1} \sum_{y \in Y} P(y) = \mathbbm{1} / n~.
\end{align}
\end{ProThe}

Now we can calculate the asymptotic symbol probabilities, using the density
matrix formalism for computing probabilities of measurement outcomes
\cite{nielsen}, and $\rho^s$.

\begin{Prop}
\label{the:asy}
A QG's symbol distribution depends only on the dimensions of the projection
operators and the Hilbert space.
\end{Prop}

\begin{ProProp}
Denote the \emph{trace operator} by $\tr$, then we have
\begin{align}
\Prob(y) 
\notag
	& = \tr \left( T^{\dagger}(y) \rho^s T(y) \right)\\
\notag
	& = n^{-1} \tr \left( T^{\dagger}(y) \mathbbm{1} T(y) \right)\\
\notag
	& = n^{-1} \tr \left( P^{\dagger}(y) U^{\dagger} U P(y) \right)\\
\notag
	& = n^{-1} \tr \left( P^{\dagger}(y) \mathbbm{1} P(y) \right)\\
\notag
	& = n^{-1} \tr \left( P(y) \right)\\
	& = n^{-1} \dimn \; P(y) ~.
\label{eqn:qprs}
\end{align}
\end{ProProp}

Although the single-symbol distribution is determined by the dimension
of the subspaces onto which the $P(y)$ project, distributions of words
$y^L$ with $L>1$ are not similarly restricted.
The asymptotic word probabilities $\Prob(y^L)$ are:
\begin{equation}
\Prob(y^L) = \tr \left( T^{\dagger}(y^L)\rho^s T(y^L) \right) ~.
\label{eqn:qprw}
\end{equation}
No further simplification is possible for the general case.

Analogous results follow for QRs, except that the calculations are suitably
modified to use $T(x)$. 

\subsection{Finitary Process Hierarchy}
\label{sec:rec-gen}

\begin{figure}  
\begin{center}
\resizebox{!}{1.50in}{\includegraphics{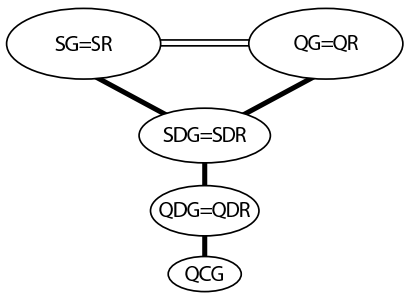}}
\end{center}
\caption{Finitary process language hierarchy: Each circle represents the set of
	process languages recognized or generated by the inscribed machine class.
	Increasing height indicates proper containment; machine classes at the same
	height are not directly comparable. The hierarchy summarizes the theorems,
	propositions, and corollaries in Secs.~\ref{sec:SRSGProp} and
	\ref{sec:rec-gen}.
  }
\label{fig:lang-I}
\end{figure}

To better appreciate what these machines are capable of we amortize the effort
in developing the preceding results to describe the similarities and
differences between quantum recognizers and generators, as well as between
stochastic and quantum automata. We collect the results, give a summary and
some interpretation, and present a road map (Fig. \ref{fig:lang-I}) that lays
out the computational hierarchy of finitary quantum processes. As above,
$\stochL(M)$ denotes the stochastic language associated with machine or
machine type $M$ and $\PRCL(C)$, the set of stochastic languages generated
or recognized by all machines in class $C$.

\begin{Prop}
QCGs are deterministic.
\end{Prop}
\begin{ProProp}
Since all projection operators have dimension one, all transition matrices
have at most one nonzero element per row. This is the condition for being a
QDG.
\end{ProProp}

Non-degenerate measurements always define a QDG. There are degenerate
measurements, however, that also can lead to QDGs, as we will show shortly.
One concludes that $\PRCL (QCG) \subset \PRCL (QDG)$.

We now show that for any QDG there is an SDG generating the same stochastic
language. Thereby we establish \emph{observational} equivalence between
the different classes of machine.

\begin{The}
\label{the:qeDG}
Every $\stochL(QDG)$ is generated by some SDG:
$\PRCL (QDG) \subseteq \PRCL (SDG)$.
\end{The}

\begin{ProThe}
We show that the SDG generating $\stochL(QDG)$ is the \emph{equivalent} SDG, as
defined in Sec.~\ref{sec:qg}, and that the QDG $\mach$ and its
\emph{equivalent} SDG $\mach^{\prime}$ generate the same word distribution and
so the same stochastic language.

The word probabilities $\Prob_\mach (y^L)$ for $\mach$ are calculated using
Eq.~(\ref{eqn:qprw}) and the QDG's transition matrices $T_{\mach^\prime}$: 
\begin{align}
\notag
\Prob_\mach(y^L)
	& = \tr \left( T_{\mach}^{\dagger}(y^L)\rho^s T_{\mach}(y^L) \right)\\
\notag
	& = n^{-1}\tr(T^{\dagger}T) \\
\notag
	& = n^{-1}\sum_i [T^{\dagger}T]_{ii}\\
\notag
	& = n^{-1}\sum_i \sum_j T_{ij}^{\dagger}T_{ji}\\
\notag
	& = n^{-1}\sum_{ij} T_{ij}^2~.
\end{align}

The word probabilities $\Prob_{M^{\prime}} (y^L)$ for $M^{\prime}$ are
calculated using Eq.~(\ref{eq:pryl}) and the SDG's transition matrices
$T_\mach$:
\begin{align}
\notag
\Prob_{\mach^{\prime}}(y^L)
 & = \braket{ \pi^0 |  T_{{\mach}^{\prime}} (y^L)  | \eta } \\
\notag
 & = \sum_{i=0}^{n-1} \left( \pi^0_i \sum_j (T_{{\mach}^{\prime}}
(y^L))_{ij}\right)\\
 & = n^{-1} \sum_{i,j=0}^{n-1}(T_{{\mach}^{\prime}} (y^L))_{ij}~.
\end{align}
Since $(T_\mach(y^L))_{ij}^2 = (T_{\mach^{\prime}} (y^L))_{ij}$, from the definition of an
\emph{equivalent} SDG, the claim follows.
\end{ProThe}

More than one QDG can be observationally equivalent to a given SDG. The 
reason for this to occur is that the quantum mechanical phases of the
transition amplitudes cancel in the transformation from a QDG.

We can now easily characterize languages produced by QDGs.
\begin{Cor}
\label{cor:QDGreg}
For every QDG, $\supp~\stochL(QDG)$ is a regular language.
\end{Cor}

\begin{ProCor}
This follows directly from Thm.~\ref{the:qeDG} and Cor.~\ref{cor:SGreg}.
\end{ProCor}

\begin{Cor}
\label{cor:QDGProc}
For every QDG, $\stochL(QDG)$ is a process language. 
\end{Cor}

\begin{ProCor}
This follows directly from Thm.~\ref{the:qeDG} and Cor.~\ref{cor:SGproc}.
\end{ProCor}

With this we can begin to compare the descriptive power of the
different machine types.

\begin{Prop}
\label{prop:QRQG}
QGs and QRs are equivalent: They recognize and generate the same set of
stochastic languages, respectively: $\PRCL(QG) = \PRCL(QR)$. 
\end{Prop}

\begin{ProProp}
Consider QG's transition matrices $T(y) = UP(y)$ and form a new set $T(x) =
UP(x)$ in which $P(x) = P(y)$, associating the QR's input $X$ with the QG's
output $Y$. The $T(x)$ define a QR that recognizes $\stochL(QG)$. It follows
that $\PRCL(QG) \subseteq \PRCL(QR)$.

Now consider QR's transition matrices $T(x) = UP(x)$ and form a new set
$T(y)$ in which $P(y) = P(x)$, associating inputs and outputs as above.
The $T(y)$ define a QG that generates $\stochL(QR)$.

It follows that $\PRCL(QG) = \PRCL(QR)$.
\end{ProProp}

\begin{Cor}
\label{cor:QDRQDG}
QDGs and QDRs are equivalent: They recognize and generate the same set of
stochastic languages, respectively:
$\PRCL(QDG) = \PRCL(QDR)$. 
\end{Cor}

\begin{ProCor}
Prop. \ref{prop:QRQG}'s proof goes through if one restricts to deterministic
machines.
\end{ProCor}

\begin{Cor}
\label{cor:QDRreg}
For every QDR, $\supp~\stochL(QDR)$ is a regular language.
\end{Cor}

\begin{ProCor}
This follows directly from Cor.~\ref{cor:SGreg} and Cor.~\ref{cor:QDRQDG}.
\end{ProCor}

\begin{Cor}
For every QDR, $\stochL(QDR)$ is a process language. 
\end{Cor}

\begin{ProCor}
This follows directly from Cor.~\ref{cor:QDGProc} and Cor.~\ref{cor:QDRQDG}.
\end{ProCor}

\begin{Prop}
\label{prop:SDGQDG}
There exists an SDG such that $\prcl(SDG)$ is not generated by any QDG.
\end{Prop}

\begin{ProProp}
The process language generated by the SDG given by
$T(0) = \left( \tfrac{1}{\sqrt{2}} \right)$ and
$T(1) = \left( 1-\tfrac{1}{\sqrt{2}} \right)$
(a biased coin) cannot be generated by any QDG. According to
Prop.~\ref{the:asy} $\Prob(y) = n^{-1} \rank P(y)$, which is a rational
number, whereas $\Prob(y)$ for the above biased coin is irrational. 
\end{ProProp}

\begin{Cor}
$\PRCL(QDG) \subset \PRCL(SDG)$.
\end{Cor}

\begin{ProCor}
From Thm.~\ref{the:qeDG} and Prop.~\ref{prop:SDGQDG}.
\end{ProCor}

\begin{Cor}
$\PRCL(QDR) \subset \PRCL(SDR)$:
\end{Cor}

\begin{ProCor}
From Cor.~\ref{cor:SDRSDG}, Cor.~\ref{cor:QDRQDG}, Thm.~\ref{the:qeDG}, and
Prop.~\ref{prop:SDGQDG}.
\end{ProCor}

At this point it is instructive to graphically summarize the relations
between recognizer and generator classes. Figure~\ref{fig:lang-I} shows
a machine hierarchy in terms of sets of languages recognized or generated.
The class of QCGs is at the lowest level. This is contained in the class of
QDGs and QDRs. The languages they generate or recognize are properly
included in the set of languages generated or recognized by classical
deterministic machines---SDGs and SDRs. These, in turn, are included in
the set of languages recognized or generated by classical nondeterministic
machines, SGs and SRs, as well as QRs and QGs.

The preceding results serve to indicate how portions of the finitary process
hierarchy are organized. However, there is still more to understand. For
example, the regularity of the support of finitary process languages,
the hierarchy's dependence on acceptance threshold $\delta$, and the
comparability of stochastic and quantum nondeterministic machines await
further investigation.


\section{Quantum Generators and Finitary Processes: Examples}

To appreciate what can be done with quantum machines, we will illustrate
various features of QTs by modeling several prototype quantum dynamical
systems. We start out with deterministic QGs, building one to model a
physical system, and end on an example that illustrates a nondeterministic QT.

\subsection{Two-State Quantum Processes}
\label{sec:tsqp}

According to Prop.~\ref{the:asy} the symbol distribution generated by a QG
only depends on the dimension of the projection operator and the dimension
of the Hilbert space. What are the consequences for two-state QGs? First
of all, according to Cor.~\ref{cor:maxA} the maximum alphabet size is $2$.
The corresponding projection operators can either have dimension $2$ (for a
single-letter alphabet) or dimension $1$ for a binary alphabet. The only
symbol probabilities possible are $\Prob(y)=1$ for the single-letter alphabet
and $\Prob(y)=1/2$ for a binary alphabet. So one can set aside the
single-letter alphabet case as too simple.

We also see that a binary-alphabet two-state QDG can produce only a
highly restricted set of process languages. It is illustrative to look
at the possible \emph{equivalent} SDGs. Their state-to-state transition
matrices are given by
\begin{equation}
T = \left(\begin{array}{cc} p & 1-p \\ 1-p & p \end{array}\right)~,
\end{equation}
with $p \in \{0,1/2,1\}$.

For $p=1/2$, for example, this is the fair coin process. It becomes
immediately clear that the Golden Mean and the Even processes, which are
modeled by two-state classical automata, cannot be represented with
a two-state QDG. (The three-state models are given below.)

\subsubsection{Iterated Beam Splitter}

Let's consider a physical two-state process and build a quantum generator
for it.

\begin{figure}  
\begin{center}
\resizebox{!}{1.25in}{\includegraphics{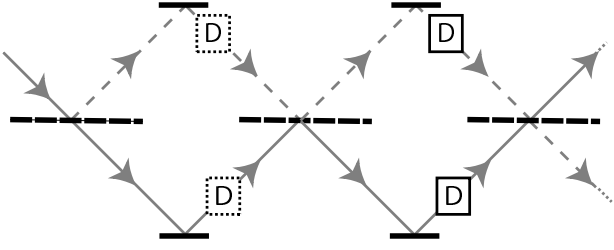}}
\end{center}
\caption{Experimental set-ups for the iterated beam splitter: Solid lines are
  mirrors; beam splitters, horizontal dashed lines. Photon nondemolition
  detectors, marked
  as D, are placed between every pair of beam splitters. Under measurement
  protocol I all detectors are in operation; under protocol II only
  the solid-line detectors are activated. The apparatus is repeated
  indefinitely to the right.
  }
\label{fig:had-exp}
\end{figure}

The \emph{iterated beam splitter} is an example that, despite its simplicity,
makes a close connection with real experiment. Figure~\ref{fig:had-exp} shows
the experimental apparatus. Photons are sent through a beam splitter (thick
dashed line), producing two possible paths. The paths are redirected by
mirrors (thick horizontal solid lines) and recombined at a second
beam-splitter. From this point on the same apparatus is repeated indefinitely
to the right. After the second beam-splitter there is a third and a fourth
and so on. Single-photon quantum nondemolition detectors are located along the paths, between
every pair of beam-splitters. One measures if the photon travels in the
upper path and the other determines if the photon follows the lower path.

This is a quantum dynamical system: a photon passing repeatedly through
various beam splitters. It has a two-dimensional state space with two
eigenstates---``above'' and ``below''. Its behavior is given by the
evolution of a state vector $\bra{\psi}$. The overall process can be represented
in terms of a unitary operation for the beam splitter and projection operators
for the detectors. The unitary operator for the beam splitter is the Hadamard
matrix $U_H$:
\begin{equation}
U_H = \frac{1}{\sqrt{2}}
	\left(\begin{array}{cc} 1 & 1 \\ 1 & -1 \end{array}\right) ~.
\label{eqn:had}
\end{equation}
The measurement operators have the following matrix representation in the
experiment's eigenbasis:
\begin{equation}
P(0) = \left(\begin{array}{cc} 1 & 0 \\ 0 & 0 \end{array}\right) ~\mathrm{and}~
P(1) = \left(\begin{array}{cc} 0 & 0 \\ 0 & 1 \end{array}\right) ~,
\label{eqn:hadP}
\end{equation}
where the measurement symbol $0$ stands for ``above'' and symbol $1$
stands for ``below''.

Before we turn to constructing a quantum finite-state generator to model this
experiment we can understand intuitively the sequence of outcomes that
results from running the experiment for long times. If entering the beam
splitter from above, the detectors record the photon in the upper or lower
path with equal probability. Once the photon is measured, though, it is in
that detector's path with probability $1$. And so it enters the beam
splitter again via only one of the two possible paths. Thus, the second
measurement outcome will have the same uncertainty as the first: the
detectors report ``above'' or ``below'' with equal probability. The
resulting sequence of outcomes after many beam splitter passages is
simply a random sequence. Call this measurement protocol I.

Now consider altering the experiment slightly by removing the detectors
after every other beam splitter. In this configuration, call it protocol II,
the photon enters the first beam splitter, does not pass a detector and
interferes with itself at the next beam splitter. That interference, as we
will confirm shortly, leads to destructive interference of one path after
the beam splitter. The photon is thus in the same path after the second
beam splitter as it was before the first beam splitter. A detector placed
after the second beam splitter therefore reports with probability $1$ that
the photon is in the upper path, if the photon was initially in the upper
path. If it was initially in the lower path, then the detector reports
that it is in the upper path with probability $0$. The resulting sequence
of upper-path detections is a very predictable sequence, compared to the
random sequence from protocol I.

\begin{figure}
\begin{center}
\resizebox{3.5in}{!}{\includegraphics{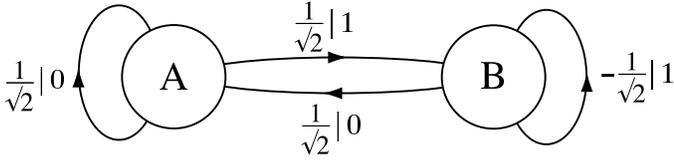}}
\end{center}
\caption{Quantum finite-state machine for the iterated beam splitter:
  The resulting symbol sequences are statistically identical to the
  sequences obtained with the measurement protocols I and II
  shown in Fig.~\ref{fig:had-exp}.
  When no measurement is made, transitions along all edges occur.
  }
\label{fig:had-qfg}
\end{figure}

We now construct a QG for the iterated-beam splitter using the matrices of
Eqs.~(\ref{eqn:had})-(\ref{eqn:hadP}) and the stationary state of
Eq.~(\ref{eq:rhos}). The output
alphabet consists of two symbols denoting detection ``above'' or ``below'': $Y
= \{0,1\}$. The set of states consists of the two eigenstates of the system
``above'' and ``below'': $Q = \{A, B\}$. The transition matrices are:
\begin{subequations}
\begin{align}
T(0) = U_H  P(0) & = \frac{1}{\sqrt{2}}
	\left(
	\begin{array}{cc}
		1 & 0 \\
		1 & 0
	\end{array}
	\right) ~,\\
T(1) = U_H  P(1) & = \frac{1}{\sqrt{2}}
	\left(
	\begin{array}{cc}
		0 & 1 \\
		0 & -1
	\end{array}
	\right) ~.
\end{align}
\end{subequations}
The resulting QG turns out to be deterministic, as can be seen from its
graphical representation, shown in Fig.~\ref{fig:had-qfg}. 

The word distribution for the process languages generated by protocols I and II
are obtained from Eq.~(\ref{eqn:qprw}). Word probabilities for protocol I
(measurement at each time step) are, to give some examples: 
\begin{subequations}
\begin{align} 
\Prob(0) & = n^{-1} \rank (P(0)) = \frac{1}{2}~,\\
\Prob(1) & = n^{-1} \rank (P(1)) = \frac{1}{2}~,\\
\Prob(00) 
	& = \tr \left( T^{\dagger}(0)T^{\dagger}(0) \rho^s T(0)T(0) \right)
	= \frac{1}{4}~,\\
\Prob(01) & = \Prob(10) = \Prob(11) = \frac{1}{4}~.
\end{align} 
\end{subequations}
Continuing the calculation for longer words shows that the word
distribution is uniform at all lengths $\Prob(y^L) = 2^{-L}$. 

For protocol II (measurement every other time step) we find:
\begin{subequations}
\begin{align} 
\Prob(0) & = \tr  \left( T^{\dagger}(\lambda 0) \rho^s T(\lambda 0) \right) 
	= \frac{1}{2}~,\\
\Prob(1) & = \tr \left( T^{\dagger}(\lambda 1) \rho^s T(\lambda 1) \right) 
	= \frac{1}{2}~,\\
\Prob(00) 
	& = \tr \left(T^{\dagger}(\lambda 0 \lambda 0) \rho^s T(\lambda 0 \lambda 0)  \right)
	= \frac{1}{2}~,\\
\Prob(11) 
	& = \tr \left(T^{\dagger}(\lambda 1 \lambda 1) \rho^s T(\lambda 1 \lambda 1) \right)
	= \frac{1}{2}~,\\
	\Prob(10) & = \Prob(01) = 0~.
\end{align} 
\end{subequations}
If we explicitly denote the output at the unmeasured time step as $\lambda$,
the sequence $11$ turns into $\lambda 1 \lambda 1$, as do the other sequences
in protocol II. As one can see, the word probabilities calculated from the QDG
agree with our earlier intuitive conclusions.

\begin{figure}  
\begin{center}
\resizebox{3.5in}{!}{\includegraphics{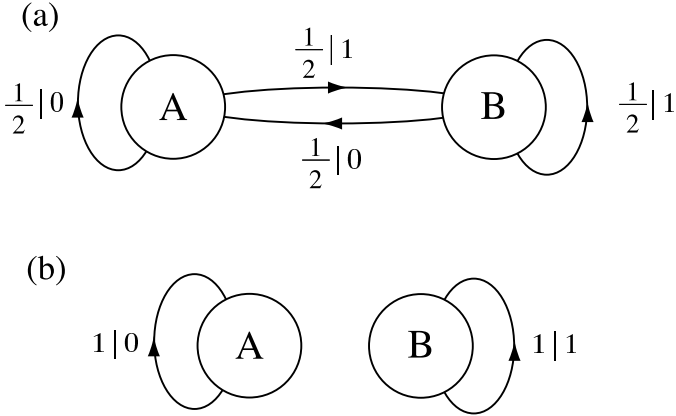}}
\end{center}
\caption{Classical deterministic generators for the iterated beam splitter:
  (a) Protocol I and (b) protocol II,
  $\mperiod = 2$.
  (Cf. Fig.~\ref{fig:had-exp}.)
  }
\label{fig:had-class}
\end{figure}

Comparing the iterated beam splitter QDG to its classically equivalent
SDG reveals several crucial differences in performance. Following the
recipe from Sec.~\ref{sec:rec-gen}, on how to build an SDG from a QDG,
gives the classical generator shown in Fig.~\ref{fig:had-class}(a). Its
transition matrices are:
\begin{equation}
T(0) = \frac{1}{2}
	\left(\begin{array}{cc} 1 & 0 \\ 1 & 0 \end{array}\right)
	~\mathrm{and}~
T(1) = \frac{1}{2}
	\left(\begin{array}{cc} 0 & 1 \\ 0 & 1 \end{array}\right) ~.
\end{equation}

The symbol sequence generated by this SDG for protocol I is the
uniform distribution for all lengths, as can be easily verified using
Eq.~(\ref{eq:pryl}) or, since it is deterministic, Eq.~(\ref{eq:DetProbyL}).
This is equivalent to the language generated by the QDG under protocol I. However, the
probability distribution of the sequences for the generator under protocol II,
ignoring every second output symbol, is still the uniform distribution for
all lengths $L$. This could not be more different from the language
generated by the QDG under protocol II.

The reason is that the classical machine is unable to capture the
interference effects present in experimental protocol II. A second SDG
has to be constructed from the QDG's transition matrices for set-up II.
This is done by carrying out the matrix product first and then forming
its equivalent SDG. The result is shown Fig.~\ref{fig:had-class}(b).
Its transition matrices are:
\begin{equation}
T(0) = \frac{1}{2}
	\left(\begin{array}{cc} 1 & 0 \\ 0 & 0 \end{array}\right) ~\mathrm{and}~
T(1) = \frac{1}{2}
	\left(\begin{array}{cc} 0 & 0 \\ 0 & 1 \end{array}\right) ~.
\end{equation}

The two classical SDGs are clearly (and necessarily) different. Thus, a
single QG can model a quantum system's dynamics for different measurement
protocols. Whereas an SG only captures the behavior of each individual
experimental set-up. This simple example serves to illustrate the utility
of QGs over SGs in modeling the behavior of quantum dynamical systems.

\subsection{Three-State Quantum Processes}

\subsubsection{Golden Mean Quantum Machine}

\begin{figure}  
\begin{center}
\resizebox{2.34in}{!}{\includegraphics{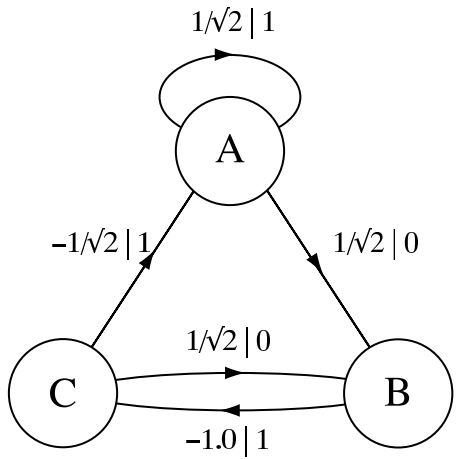}}
\end{center}
\caption{\label{fig:qgldm}Quantum generator for the Golden Mean Process.
  }
\end{figure}

Recall the classical Golden Mean generator of
Sec.~\ref{sec:StochasticGenerators}. A QDG,
which generates the same process language, is shown in Fig.~\ref{fig:qgldm}.
Consider a spin-$1$ particle subject to
a magnetic field that rotates its spin. The state evolution can be described
by the unitary matrix
\begin{equation}
U = \left(
	\begin{array}{ccc}
		\frac{1}{\sqrt{2}} & \frac{1}{\sqrt{2}} & 0
		\\ 0 & 0 & -1
		\\ -\frac{1}{\sqrt{2}} & \frac{1}{\sqrt{2}} & 0 
	\end{array}
	\right) ~,\\
\label{eqn:qgldm}
\end{equation}
which is a rotation in $\mathbb{R}^3$ around the y-axis by angle
$\tfrac{\pi}{4}$ followed by a rotation around the x-axis by $\tfrac{\pi}{2}$.

Using a suitable representation of the spin operators $J_i$ \cite[p.199]{peres},
such as:
$
J_x = \left(
	\begin{smallmatrix}
        0 & 0 & 0 \\
        0 & 0 & i \\
        0 & -i & 0
	\end{smallmatrix}
    \right)
$,
$
J_y = \left(
	\begin{smallmatrix}
        0 & 0 & i \\
        0 & 0 & 0 \\
        -i & 0 & 0
	\end{smallmatrix}
    \right)
$, and
$
J_z = \left(
	\begin{smallmatrix}
        0 & i & 0 \\
        -i & 0 & 0 \\
        0 & 0 & 0
	\end{smallmatrix}
    \right) ,
$
the relation $P_i = 1 - J_i^2$
defines a one-to-one correspondence between the projector $P_i$ and the
square of the spin component along the $i$-axis. This measurement
poses the yes-no question, Is the square of the spin component along
the $i$-axis zero? Consider measuring $J_y^2$. Then $U$, the projection
operator $P(0) = \ket{100}\bra{100} + \ket{001}\bra{001}$ for $y$-component
zero, and that for nonzero $y$-component $P(1) = \ket{010}\bra{010}$,
define a quantum generator whose outputs are a sequence of the
spin's $y$-component.

The transition matrices $T(y)$ are then
\begin{subequations}
\begin{align}
T(0) & = U P(0) =
	\left(
	\begin{array}{ccc}
		0 & \frac{1}{\sqrt{2}} & 0  \\
		0 & 0 & 0 \\
		0 & \frac{1}{\sqrt{2}} & 0
	\end{array}
	\right) ~, \\
T(1) & = U P(1)  =
	\left(
	\begin{array}{ccc}
		\frac{1}{\sqrt{2}} & 0 & 0
		\\ 0 & 0 & -1
		\\ -\frac{1}{\sqrt{2}} & 0 & 0 
	\end{array}
	\right) ~.
\end{align}
\end{subequations}

To illustrate that this QDG produces the Golden Mean word distribution
we show how to calculate several of the word probabilities using
Thm.~\ref{the:asy} and Eq.~(\ref{eqn:qprw}): 
\begin{subequations}
\label{eqn:qgm}
\begin{align} 
\Prob(0) & =  n^{-1} \rank (P(0)) = \frac{1}{3}~,\\
\notag
\Prob(1) & =  n^{-1} \rank (P(1)) = \frac{2}{3}~,\\
\Prob(011) & = \tr \left(T^{\dagger}(011) \rho^s T(011) \right) = \frac{1}{6}. 
\end{align} 
\end{subequations}

\subsubsection{Quantum Even Process}

\begin{figure}  
\begin{center}
\resizebox{2.34in}{!}{\includegraphics{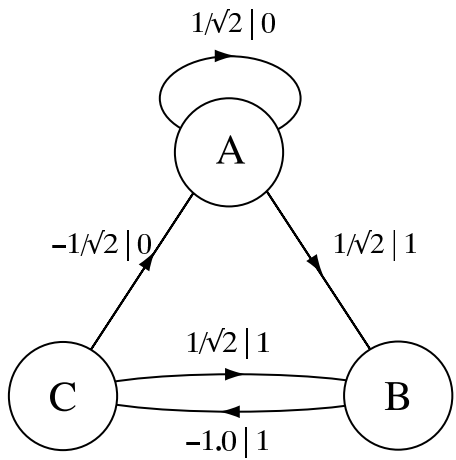}}
\end{center}
\caption{Quantum generator for the Even Process.
  }
\label{fig:qeven}
\end{figure}

\begin{figure}  
\begin{center}
\resizebox{!}{2.50in}{\includegraphics{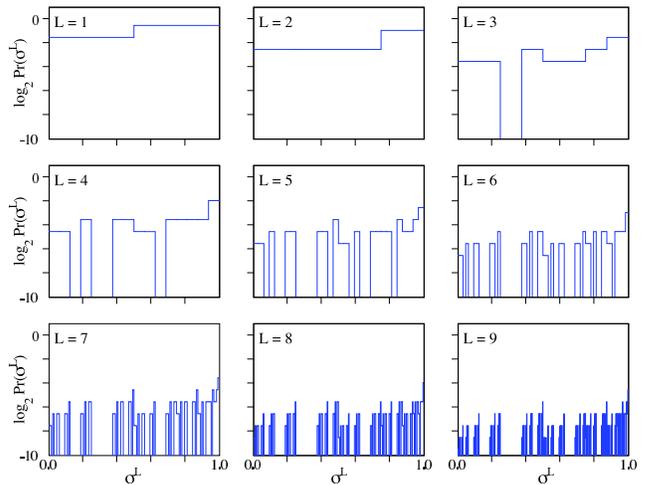}}
\end{center}
\caption{
  Process language of the Even QDG.
  }
\label{fig:evdistr}
\end{figure}

The next example is a quantum representation of the Even Process. Consider the
same spin-$1$ particle. This time the $J_x^2$ component is chosen as
observable. Then $U$ and $P(0) = \ket{100}\bra{100}$ and $P(1) =
\ket{011}\bra{011}$ define a quantum finite-state generator. The QDG is shown
in Fig.~\ref{fig:qeven}. The word distributions for lengths up to $L = 9$ are
shown in Fig.~\ref{fig:evdistr}.

Note that the unitary evolution for the Golden Mean Process and the Even
Process are the same, just as the state-to-state transition matrices were
the same for their classical versions. The partitioning into subspaces induced
by the projection operators leads to the (substantial) differences in the
word distributions; cf. Figs. \ref{fig:stol} versus \ref{fig:evdistr}.

The dependence on subspace partitioning indicates a way to count the number of
QGs for each unitary evolution $U$. For 3-dimensional Hilbert spaces this is
rather straightforward. For each unitary matrix and with a binary alphabet
we have three choices for partitioning subspaces of the Hilbert space: one
subspace is two-dimensional and the others one-dimensional. This yields three
QGs that are distinct up to symbol exchange ($0\leftrightarrow1$). For the
unitary matrix that generates the Golden Mean and the Even Process
(Eq.~(\ref{eqn:qgldm})) the third QG turns out to be nondeterministic. But no
phase interference is possible and it generates the Golden Mean process
language. The potential non-Markovian (sofic) nature of these quantum
processes has been discussed in Ref.~\cite{Wies07a}.

This very limited number of possible QGs for any given unitary matrix is
yet another indication of the limitations of QGs. Classical SGs do not have
the same structural restrictions, since they are not bound by orthogonal
partitioning into subspaces, for example. The saving grace for QGs is that
they have complex transition amplitudes and so can compute with phase, as long
as they are not observed. This is reflected in the distinct languages
generated by one QG under different measurement protocols
\cite{Wies06d}.

\subsection{Four-State Quantum Process}

We are now in a position to explore the full capabilities of QTs, turning
from generators to transducers. The
following example illustrates quantum machines by using the tools required
to investigate information processing of quantum dynamical systems.

\subsubsection{Quantum Transducer for Trapped Ions}

Consider an atom exposed to short wavelength radiation---the core of
numerous experiments that investigate electronic structure and dynamics.
The usual procedure is a one-time experiment, exposing the atom to
radiation and monitoring changes in structure through electron or photon
detectors. As a particular set-up we choose ion-trap experiments found
in low-temperature physics and quantum computation implementations, as 
described in Ref. \cite{galindo:02}. For our present purposes it will
be sufficient to review the general physical setting.

\begin{figure}
\begin{center}
\resizebox{!}{1.25in}{\includegraphics{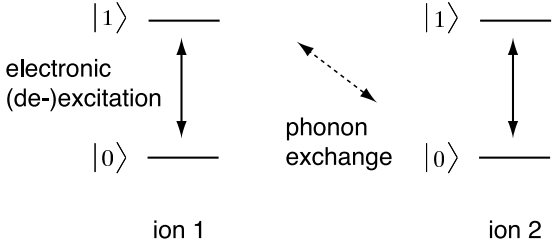}}
\end{center}
\caption{Schematic view of two vibrationally-coupled trapped ions undergoing
  electronic excitation. Only the two electronic levels of interest are drawn.
  }
\label{fig:ionstp}
\end{figure}

Imagine a pair of ions kept in a trap by laser fields and static
electromagnetic fields. Only two of the electronic levels of each ion are
of interest: the ground state and an excited state. Call these level $0$
and level $1$, respectively. A third auxiliary level is required for laser
cooling and other operations, which we leave aside here since it has no
significance for the description of the process. The two ions are coupled
to each other through phonon exchange, as shown schematically in
Fig.~\ref{fig:ionstp}.

By choosing suitable wavelengths several distinct operators can be
implemented. One of them is a Hadamard operator that produces a
superposition of electronic states $\ket{0}$ and $\ket{1}$. Another is a phase
operator that yields an entangled state of the two ions. The respective laser
pulses, so-called \emph{Rabi} pulses, induce an electronic excitation and a
vibrational excitation. The result is vibrational coupling of the four levels.
All other operations are combinations of these two; see
Ref.~\cite{galindo:02}. The operators are named $U_a$, $U_b$, and $U_c$;
matrix representations are given below. As is already familiar from the
iterated beam splitter, the operators are activated repeatedly one after
the other in a closed loop and as such constitute a quantum dynamical system.

To model the quantum dynamical system the state vector and operator matrices
need to be specified. The four basis states spanning the Hilbert space
are given by:
\begin{align*}
\bra{\phi_A} & = \bra{1000},~
\bra{\phi_B} & = \bra{0100},~
\bra{\phi_C} & = \bra{0010},\\
\bra{\phi_D} & = \bra{0001}.
\end{align*}
$\phi_A$ corresponds to both ions in electronic state $\ket{0}$. $\phi_B$
corresponds to ion $1$ in state $\ket{0}$ and ion $2$ in state $\ket{1}$,
and so on. The three unitary operations in matrix form are:
\begin{subequations}
\begin{align}
U_a & = H \otimes H = 
\frac{1}{2} \left( \begin{array}{rrrr} 1 & 1 & 1 & 1 \\ 
										1 & -1 & 1 & -1 \\
										1 & 1 & -1 & -1 \\
										1 & -1 & -1 & 1 
\end{array}\right) ~,\\
U_b & = 
\left( \begin{array}{rrrr}  1 & 0 & 0 & 0 \\ 
							0 & 1 & 0 & 0 \\
							0 & 0 & -1 & 0 \\
							0 & 0 & 0 & -1
\end{array}\right) ~,\\
U_c & = H \otimes I = 
\frac{1}{\sqrt{2}} \left( \begin{array}{rrrr}
		1 & 0 & 1 & 0 \\ 
		0 & 1 & 0 & 1 \\
		1 & 0 & -1 & 0 \\
		0 & 1 & 0 & -1 
\end{array}\right) ~.
\end{align}
\end{subequations}

\begin{figure*}
\begin{center}
\resizebox{3.94in}{!}{\includegraphics{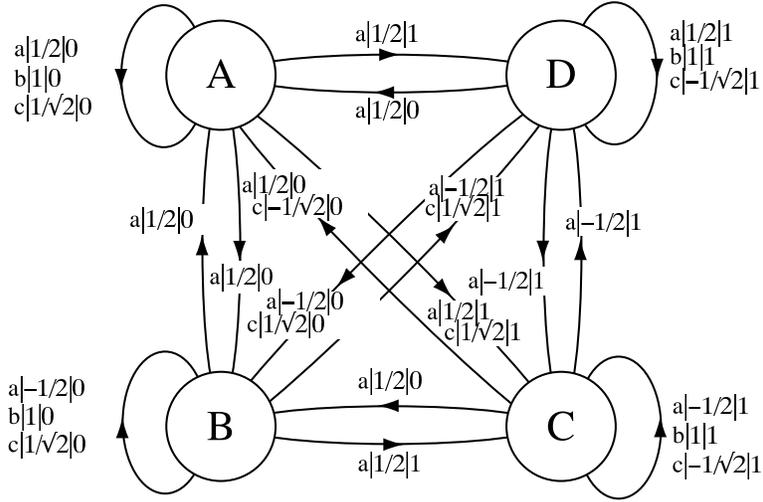}}
\end{center}
\caption{Quantum transducer for a trapped-ion system exposed to
  radiation of various wavelengths. The input alphabet $X = \{a,b,c\}$ and
  output alphabet $Y = \{0,1\}$ represent unitary operations and electronic
  states, respectively.
  }
\label{fig:deu-qm}
\end{figure*}

The projection operators are chosen to measure the electronic state of ion
$1$ only and have the matrix form:
\begin{align}
P(0) = \left(
	\begin{array}{cccc}
		1 & 0 & 0 & 0 \\
		0 & 1 & 0 & 0 \\
        0 & 0 & 0 & 0 \\
		0 & 0 & 0 & 0 
	\end{array}
	\right) \mathrm{and}~
P(1) = \left(
	\begin{array}{cccc}
		0 & 0 & 0 & 0 \\
		0 & 0 & 0 & 0 \\
		0 & 0 & 1 & 0 \\
		0 & 0 & 0 & 1 
	\end{array}
	\right).
\end{align}
The QT is now easily assembled. The set of states and the input and output
alphabets are, respectively:
$Q = \{A,B,C,D\}$, $X = \{a,b,c\}$, and $Y = \{0, 1\}$.
This QT's graph is shown in Fig.~\ref{fig:deu-qm}.

To illustrate its operation we consider two measurement protocols. For each
we use input sequence $(abc)^+$.
\begin{itemize}
\setlength{\topsep}{0pt}
\setlength{\itemsep}{0pt}
\item Measurement protocol I: Measure ion $1$ after each
unitary operation. The resulting state vector evolution is:
\begin{subequations}
	\begin{align}
		\bra{\psi_{t+1}} & = \bra{\psi_t} U_a P(y) ~,\\
		\bra{\psi_{t+2}} & = \bra{\psi_{t+1}} U_b P(y) ~,\\
		\bra{\psi_{t+3}} & = \bra{\psi_{t+2}} U_c P(y) ~.
	\end{align}
\end{subequations}
\item Measurement protocol II: Measure ion $1$ only after
three unitary operations. This leads to evolution according to
	\begin{equation}
		\bra{\psi_{t+3}} = \bra{\psi_t} U_a U_b U_c P(y) ~.
	\end{equation}
\end{itemize}

\begin{figure}  
\begin{center}
\resizebox{!}{2.50in}{\includegraphics{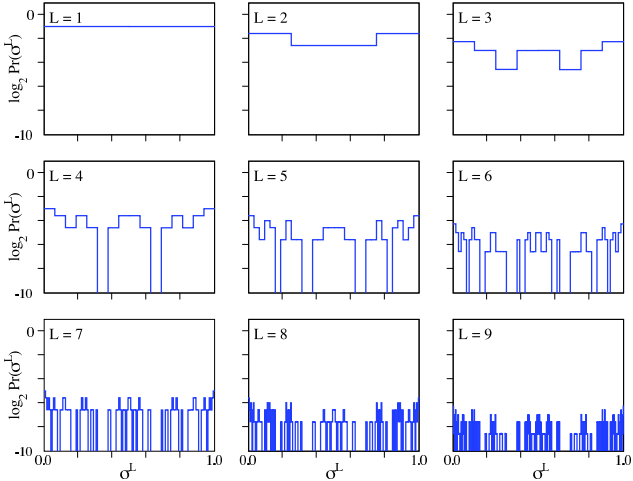}}
\end{center}
\caption{Process language generated by the trapped-ion quantum dynamical system
  of Fig.~\ref{fig:ionstp} for protocol I (measurements performed at each time
  step).
  }
\label{fig:dlp-1}
\end{figure}

\begin{figure}  
\begin{center}
\resizebox{!}{2.50in}{\includegraphics{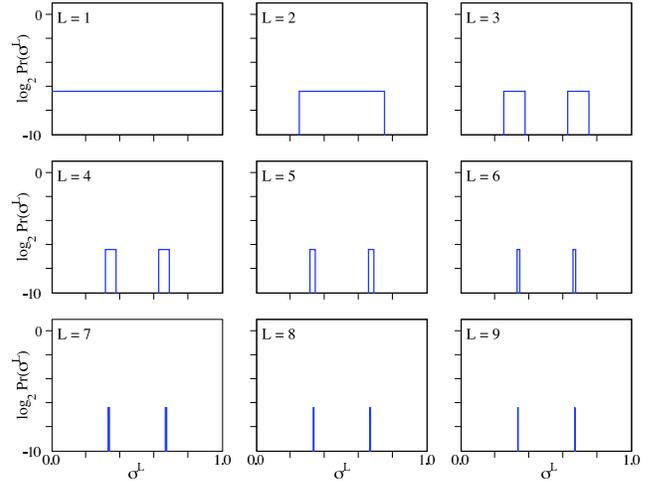}}
\end{center}
\caption{The generated process languages of the trapped-ion dynamical system
  from Fig.~\ref{fig:ionstp} for measurements performed every three time steps.
  }
\label{fig:dlp-2}
\end{figure}

The probability distributions of the observed sequences are shown in
Figs.~\ref{fig:dlp-1} and \ref{fig:dlp-2}. The two distributions differ
substantially. On the one hand, protocol II simply yields the process
language of alternating $0$s and $1$s. Protocol I, on the other hand,
yields a much larger set of allowed words. In particular, it is striking
that $\supp~\prcl^{\mathrm{II}}$ is forbidden behavior under protocol I.
The words $0101$ and $1010$ are forbidden under protocol I, whereas they
are the only allowed words of length $L=4$ under protocol II.

Not only does this example illustrate that a simple change in measurement
protocol leads to a substantial change in the observed dynamics. It is also
not clear a priori when a more complicated behavior is to be expected.
That is, more frequent measurement yields more complicated behavior.
Without quantifying how complex that complicated behavior is, it turns out
that it is not always the longer period of coherent, unperturbed unitary
evolution that yields more complex processes. This will have consequences
for feasible implementations of quantum computational algorithms. For a
quantitative discussion of the languages generated by quantum processes see
Ref.~\cite{Crut06a}.

\subsubsection{Deutsch Algorithm as a Special Case}

It turns out that the trapped-ion experiment implements a quantum algorithm
first introduced by Deutsch \cite{deutsch:85}. The algorithm provided an
explicit example of how a quantum machine could be superior to a classical one.

Consider a binary-valued function $f: \{1,2,\dots,2N\}\rightarrow\{0,1\}$.
Let $U$ be the device that computes  the function $f$. If we successively
apply $f$ to $1,2,\dots,2N$, we get a string $x^{2N}$ of length $2N$.
The problem then is to find a true statement about $x^{2N}$ by testing the
following two properties:
\begin{itemize}
\setlength{\topsep}{0pt}
\setlength{\itemsep}{0pt}
\item[A:] $f$ is not constant: Not only $0$s or only $1$s in $x^{2N}$.
\item[B:] $f$ is not balanced: Not as many $0$s as $1$s in $x^{2N}$.
\end{itemize}
If statement A is false, we can be certain that statement B is true and vice
versa. Note that both statements can be true in which case the algorithm does
not reveal anything about $f$. Deutsch and Josza \cite{deutsch:92} showed that
a quantum computer can 
determine the true statement, either A or B, after only two invocations of the
operation $U$, whereas a classical computer requires $N+1$ calls in the worst
case. Taking into account the computational steps for establishing the start
state and reading out the result, a quantum computer can evaluate the function
$f$ in constant time, whereas  a classical computer needs a time linear in
$N$.

\begin{figure}
\begin{center}
\resizebox{!}{1.25in}{\includegraphics{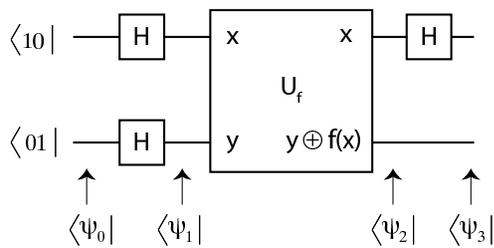}}
\end{center}
\caption{Deutsch algorithm to classify balanced and constant functions
  ($N = 2$) depicted as a quantum circuit.
  }
\label{fig:deu}
\end{figure}

To compare the algorithm with the trapped-ion dynamical system, and to keep
issues simple but still informative, we use the basic version ($N = 2$) of
the Deutsch algorithm of Ref. \cite[p. 32]{nielsen}. (Recall that in
our notation $\bra{\psi}$ is the state vector, not $\ket{\psi}$, as is
common elsewhere.) Figure~\ref{fig:deu} shows the algorithm as a quantum
circuit. Each qubit represents one ion and occupies one horizontal line. The
applied unitary transformations are shown as boxes. The overall procedure is
summarized in Table~\ref{tab:deu}. The unitary operations $H$ and $U_f$ in
Fig.~\ref{fig:deu} are the same as $H$ and $U_b$ in the trapped-ion
experiment. The unitary operator is that for a balanced function.  

The implementation of the Deutsch algorithm is equivalent to the trapped-ion
system under measurement protocol II, with $U_b$ chosen accordingly.
Measuring ion $1$ after three time steps delivers the desired answer as
output ($0$=A or $1$=B). Thus, the Deutsch algorithm
corresponds to the trapped-ion system running for three time steps.

The Deutsch algorithm task is solved with a considerable speed-up compared
to a classical implementation. Our approach is an extension of this that
focuses on what type of computation is carried out intrinsically by the
system under continuous external driving and observation. The answer is found
in the language diagrams in Figs.~\ref{fig:dlp-1} and \ref{fig:dlp-2}. Comparing these
two different views of quantum information manipulation---designed quantum
computing versus quantum intrinsic computation---suggests that the analysis
of NMR experiments with single atoms or molecules in terms of quantum
finite-state machines will be a straightforward extensions of the
preceding analysis of the Deutsch algorithm.

\begin{table}
\begin{tabular}{| l | l |}
\hline
1. Two qubits put in states & $\bra{\psi^0} = \bra{0100}$ \\
   $~~~~\bra{0}$ and $\bra{1}$, respectively. & \\
2. Hadamard transform applied & $\bra{\psi_1} = \bra{\psi^0}(H \otimes H)$\\
   $~~~~$to both qubits. & \\
3. Operation $U_f$ implementing & $\bra{\psi_2} = (-1)^{f(x)}\bra{\psi_1}(I\otimes I)$\\
   $~~~~$the function $f(x)$ is applied. & \\
4. Hadamard transform applied & $\bra{\psi_3} = \bra{\psi_2}(H\otimes I)$\\
   $~~~~$to the first qubit. & \\
5. First qubit is measured.  & $\bra{\psi_3} P(0)$\\
\hline
\end{tabular}
\caption{Deutsch algorithm to determine if $f(x)$ is balanced or constant.
  $H$ and $I$ are the Hadamard and identity matrices, respectively.
  $\otimes$ denotes the tensor product.}
\label{tab:deu}
\end{table}

\section{Concluding Remarks}

We developed a line of inquiry complementary to both quantum computation and
quantum dynamical systems by investigating intrinsic computation in quantum
processes. Laying the foundations for a computational perspective on quantum
dynamical systems, we introduced quantum finite-state transducers. Residing at
the lowest level of a quantum computational hierarchy, it is the most general
representation of a finitary quantum process. It allows for a quantitative
description of intrinsic computation in quantum processes---in terms of the
number of internal states and allowed transitions and the process language it
generates. As far as we are aware, this has not been developed before in the
quantum setting.

We laid out the mathematical foundations of these models and developed a
hierarchy of classical (stochastic) and quantum machines in terms of the set
of process languages they recognize or generate. In many cases it turned out
that quantum devices were less powerful than their classical analogs.
We saw that the limitations of quantum finite-state machines originate in
the unitarity of the transition matrices. This suggested that QTs, being
reversible,  are less powerful than nonreversible classical automata,
since the reversibility constrains the transition matrices.

However, one must be careful to not over-interpret this state of affairs.
It has been known for some time that any universal computation can be
implemented in a reversible device \cite{bennett:73}. Typically, this
requires substantially more resources, largely to store outcomes of
intermediate steps. In short, reversibility does not, in general, imply
less power for classical computers. At the end of the day, computational
resources
are variables that trade-off against each other. The 2-state QDG examples
of the Beam Splitter process illustrated such a trade-off.
Although the QDG needs more states than the equivalent SDG to generate
the same process language, different measurement protocols yielded a
new set of process languages---an aspect that makes QDGs more powerful
than SDGs \cite{Wies06d}.

These results were then applied to physical systems that could be analyzed in
terms of the process languages they generate. One example, that of two trapped
ions exhibited a process language with rich structure. This, and the fact that
the system implements a quantum algorithm, opens up a way to an
information-theoretic analysis of quantum processes. One can begin to analyze
quantum algorithms in terms of their information processing power and do so
independent of particular physical implementations. 

In particular, we have used quantum machines to define a measure of intrinsic
computation for quantum dynamical systems \cite{Crut06a,Wies07a}. The basic
questions one asks about a dynamical system's intrinsic computation---amount
of historical information stored, storage architecture, and transformations
of stored information---can now be posed for quantum systems.

Furthermore, we are developing an extension of quantum machines that supports
more general types of measurement. The resulting quantum transducers are
expected to have greater power than the current versions, possibly even greater
than stochastic transducers. Generally, we hope that ways to integrate quantum
computation and quantum dynamics will receive further attention.


\begin{acknowledgments}
The authors thank C. Ellison, D. Feldman, J. Goodwin, J. Mahoney, I. Rumanov,
M. S\'anchez-Monta\~n\'es, and C. Strelioff for helpful discussions. This
work was supported at the Santa Fe Institute and UCD under the Networks
Dynamics Program funded by the Intel Corporation and under the Computation,
Dynamics and Inference Program. Direct support was provided by DARPA
Agreement F30602-00-2-0583. KW's visit to SFI was partially supported by
an SFI Steinmetz Fellowship. KW's postdoctoral fellowship was provided by
the Wenner-Gren Foundations, Stockholm, Sweden.
\end{acknowledgments}


\bibliography{ref}

\end{document}